\documentclass[reprint,
 amsmath,amssymb,
 aps,
 pre,
]{revtex4-1}

\usepackage{natbib}
\usepackage{graphicx}
\usepackage{dcolumn}
\usepackage{bm}
\usepackage{siunitx}
\usepackage[section]{placeins} 

\begin{document}

\preprint{APS/123-QED}

\title{Pressure evolution and deformation of confined granular media during pneumatic fracturing}

\author{Fredrik K. Eriksen}
\altaffiliation[Also affiliated with ]{PoreLab, Department of Physics, University of Oslo, P.O. Box 1074 Blindern, 0316 Oslo, Norway}
\email[$~$\\$~$\\email: ]{f.k.eriksen@fys.uio.no}
\author{Renaud Toussaint}
\altaffiliation[Also affiliated with ]{PoreLab, Department of Physics, University of Oslo, P.O. Box 1074 Blindern, 0316 Oslo, Norway}
\email[$~$\\$~$\\email: ]{f.k.eriksen@fys.uio.no}
\author{Antoine L\'{e}o Turquet}
\affiliation{%
Institut de Physique du Globe de Strasbourg, Universit\'{e} de Strasbourg/EOST, Centre National de la Recherche Scientifique, 67084 Strasbourg, France\\
}%

\author{Knut J. M{\aa}l{\o}y}
\author{Eirik G. Flekk{\o}y}
\affiliation{%
PoreLab, Department of Physics, University of Oslo, P.O. Box 1074 Blindern, 0316 Oslo, Norway\\%
}%
\date{\today}

\begin{abstract}
By means of digital image correlation, we experimentally characterize the deformation of a dry granular medium confined inside a Hele-Shaw cell due to air injection at a constant overpressure high enough to deform it (from 50 to 250 kPa). Air injection at these overpressures leads to the formation of so called pneumatic fractures, i.e. channels empty of beads, and we discuss the typical deformations of the medium surrounding these structures. In addition we simulate the diffusion of the fluid overpressure into the medium, comparing it with the Laplacian solution over time, and relating pressure gradients with corresponding granular displacements. In the compacting medium we show that the diffusing pressure field becomes similar to the Laplace solution on the order of a characteristic time given by properties of the pore fluid, granular medium and the system size. However, before the diffusing pressure approaches the Laplace solution on system scale, we find that it resembles the Laplacian field near the channels, with the highest pressure gradients on the most advanced channel tips and a screened pressure gradient behind them. We show that the granular displacements more or less always move in the direction against the local pressure gradients, and when comparing granular velocities with pressure gradients in the zone ahead of channels, we observe a Bingham type of rheology for the granular paste (the mix of air and beads), with an effective viscosity $\mu_B$ and displacement thresholds $\nabla P_c$ evolving during mobilization and compaction of the medium. Such a rheology, with disorder in the displacement thresholds, could be responsible for placing the pattern growth at moderate injection pressures in a universality class like the Dielectric Breakdown Model with $\eta=2$, where fractal dimensions are found between 1.5 and 1.6 for the patterns.

\begin{description}
\item[PACS numbers]
83.60.Wc, 81.05.Rm, 47.20.Ma

\end{description}
\end{abstract}

\maketitle


\section{\label{sec:level1}Introduction}


Several processes in engineering, industry and earth sciences involve pneumatic (gas) or hydraulic (liquid) fracturing of the soil, which occurs when fluids in the ground are driven to high enough pressures to deform, fracture and generate porosity in the surrounding soil or rock. For example in environmental engineering, pneumatic or hydraulic fracturing is done to enhance the removal of hazardous contaminants in the vadose zone (soil remediation) \cite{Suth1999,Good1994}, for soil stabilization injection to ensure a solid foundation for structures \cite{Kaze2010}, or in packer tests for project planning, risk assessment and safe construction of dams and tunnels \cite{Moay2012}. In industry, hydraulic fracturing is done to enhance oil and gas recovery \cite{Mont2010,DOE2004,Naik2003}, CO$_2$ sequestration \cite{Johns2009}, water well- and geothermal energy production \cite{Will1980,Rumm1983,Clar2012}. The evolution of faults and fractures at crustal scale can also be affected by fluid flow \cite{Ghan2013,Vass2014,Ghan2015} as well as the rheology of fluid saturated faults \cite{Gore2010,Gore2013,Aoch2013}. Related natural processes such as subsurface sediment mobilization are studied in earth sciences, where seepage channels initiate due to erosion of granular soils by the fluid seeping through \cite{Berh2012}, leading later to channel and river network formations \cite{Abra2009} - physicists are interested in the interplay between fluid flow evolution and erosion patterns \cite{Kudr2016}. Also, sand injectites, mud diapirs and mud volcanoes are formed due to pore-fluid overpressure \cite{Hurs2003,Lose2003,VanR2003,Talu2003,Pral2003,Devi2003}. For example, the Lusi mud volcano in Indonesia is the biggest and most damaging mud volcano in the world \cite{Nuwe2015}, having displaced 40 000 people from their homes, and has been active since May 2006. There is an ongoing debate about how it was triggered, i.e. whether it formed naturally by an earthquake or geothermal process \cite{Mazz2007,Mazz2013,Mazz2012,Lupi2013}, or that it is a man-made consequence of a nearby drilling operation by a company probing for natural gas \cite{Ting2015}.


Fluid injection into granular media has been extensively studied in laboratory experiments and simulations, where a common method to simplify the problem is to confine the experiment within a quasi-2-dimensional geometry, i.e. a Hele-Shaw cell. In \cite{John2006,John2008b}, the decompaction, fluidization regimes, and coupling between air and granular flow were studied in dry granular media in open circular and rectangular cells during air injection at different overpressures. Similar behavior was seen for the injection of liquid into a granular material initially saturated with the same liquid \cite{John2008a}, so it is reasonable to assume that studies of pneumatic fracturing also have applications in hydraulic fracturing. The patterns formed during fluid injection into a granular medium, and evolution of the fluid-solid interface, have been found to resemble Diffusion Limited Aggregation patterns (DLA) and viscous fingering \cite{Chen2008}, a fingering instability that occurs when a less viscous liquid is injected into a porous medium containing a more viscous liquid with which it cannot mix \cite{Saff1958}. As mentioned in \cite{John2008b}, the main difference between the viscous- and granular fingering instabilities is the absence of interfacial tension in the granular case. For example, the stabilizing forces in viscous fingering are surface forces, while in granular fingering it is the build-up of friction between particles and against the confining walls. However, both instabilities are driven by the pressure gradient across the defending medium, which is largest on the longest finger tips, making more advanced fingers grow at the expense of the less advanced ones. A notable difference between air injection into a dry granular medium and a liquid saturated one is that the overpressure initially diffuses into the packing in the dry (saturated with compressible air) case, while it is already a steady-state Laplace field over the defending liquid in the saturated case.

Further, during air injection into liquid saturated granular media and suspensions, the characteristics of emerging patterns and behavior of the media depend on the injection rate, and the competition between mobilized friction and surface forces \cite{Erik2015,Sand2011,Holt2012,Kong2010,Chev2008,Alme2015a,Alme2015b,Mark2015,Alme2016,Mour2015,Wilk1983,Lovo2004,Lovo2011,Tall2009}. For example, one observes flow regimes such as two phase flow in rigid porous media \cite{Mour2015,Wilk1983,Lovo2004,Lovo2011,Tall2009}, capillary fracturing, stick-slip bubbles and labyrinth patterns \cite{Erik2015,Sand2011,Holt2012,Kong2010,Chev2008,Alme2015a,Alme2015b,Mark2015,Alme2016}. In the opposite case, during liquid injection into dry granular media \cite{Huan2012}, the flow behavior goes from stable invasion towards granular fingering for increasing flow rate and viscosity of the invading fluid. At intermediate conditions, fractures open up inside the invaded region. The same trend is shown in numerical studies for gas injection into granular media containing the same gas \cite{Nieb2012a}.


Typically, in all processes involving fluid injection into granular media, there are flow regimes where the medium has either solid-like behavior or fluid-like behavior. This is one of the special properties of granular media, which also have gas-like behavior in some cases \cite{Jaeg1996}.


In this paper, we present an experimental study on deformations and evolution of pressure fields during air injection into confined granular media. More specifically, we inject air at constant overpressure into a dry granular medium inside a Hele-Shaw cell, where air escapes at the outlet while beads cannot. The motivation of this setup is to characterize the evolution of the interstitial pressure as well as deformations surrounding pneumatic fractures in compacting granular media, and the coupling between compaction and flow. As opposed to similar experiments with open outer boundary conditions \cite{John2006,John2008a,John2008b}, here, after the flow compacts the medium there is no decompaction. We thus observe the material behavior (at high enough overpressure to displace beads) to have a transition from fluid-like to solid-like during experiments, and that eventual invasion patterns will initially resemble viscous fingering in the fluid-like regime, crossing over to stick-slip fracture propagation as the medium becomes more solid-like, until it reaches a final structure as the compacted medium has reached a completely solid-like behavior. A similar, but smaller system has been studied in numerical simulations by Niebling et al. \cite{Nieb2012a,Nieb2012b}. 
By varying the interstitial fluid viscosity, two flow regimes were identified; one with finely dispersing bubbles and large scale collective motion of particles, the other one with build-up of a compaction front and fracturing. These flow regimes depend, respectively, on whether the particles are primarily accelerated by the imposed pressure gradient in the fluid, or interactions through particle contacts. This in turn depends on the diffusivity of the interstitial fluid pressure in the granular medium. 
Due to the confined nature of our experiment, it is thought to be a laboratory analog to pneumatic and hydraulic fracturing of tight rock reservoirs where the free boundary at the surface is very distant from the injection zone. In other words, in a situation where the fractures stop before reaching a free surface such that the surrounding medium is not decompacted. Therefore, new insight into this problem may have industrial applications in addition to increase the understanding of flow and transformations in porous media.

It is also worth to mention a closely related project \cite{Turk2015}, where acoustic emissions recorded during the experiments are analyzed. There, it is shown that different stages of the invasion process can be identified acoustically in terms of characteristic frequencies and distinct microseismic events. In this paper we characterize the deformations that are the source of these emissions.


\section{\label{sec:level1}Methods}

\subsection{\label{sec:level2}Experimental setup}

\begin{figure}
\includegraphics[width=\columnwidth,height=0.9\textheight,keepaspectratio]{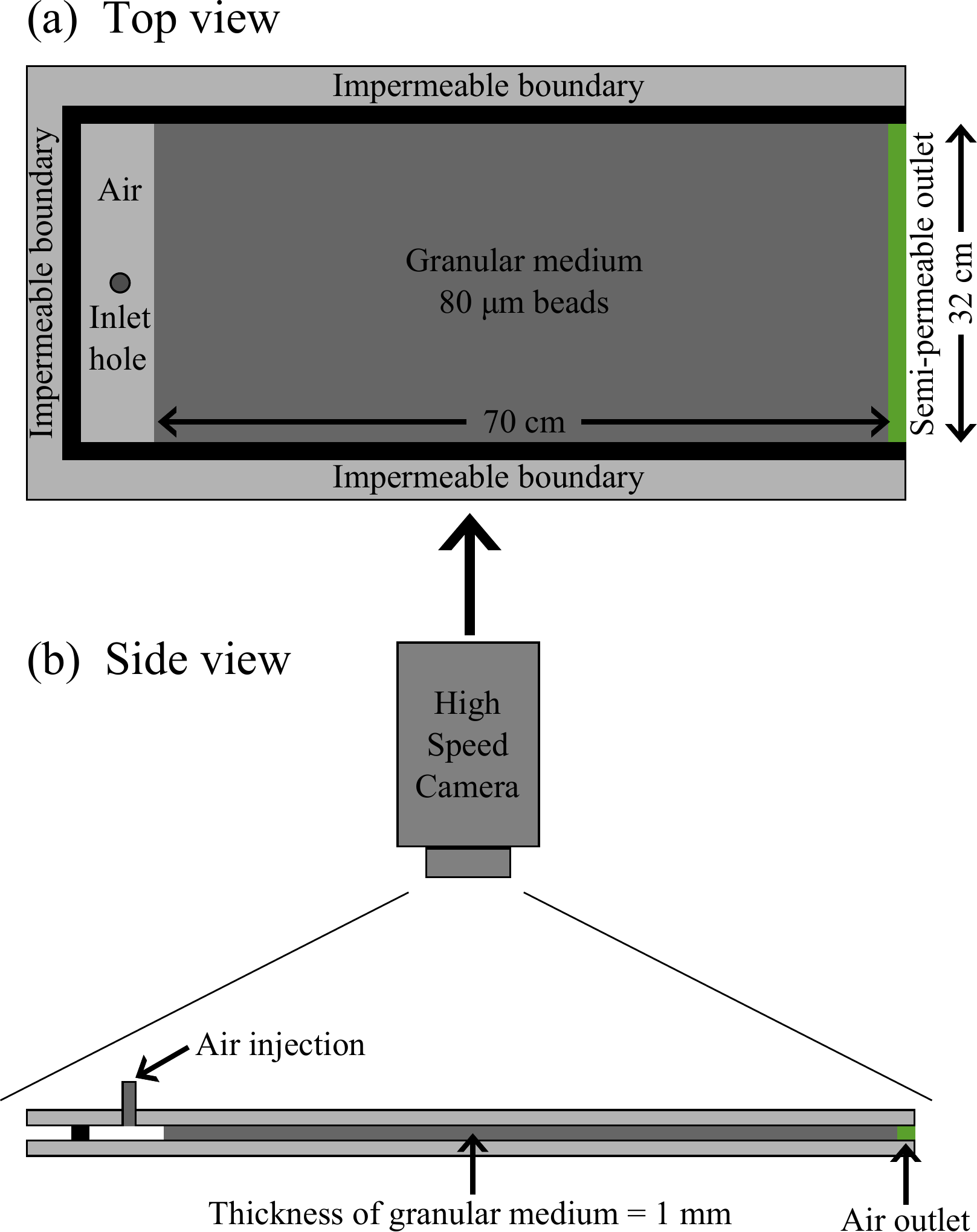}
\caption{Sketch of the experimental setup where the top-down view of the prepared cell (a) shows that the granular
medium is confined inside the cell by three impermeable boundaries and an air-permeable boundary on the outlet side. The granular medium is placed against the outlet boundary such that it has a linear interface against a region empty of beads on the sealed inlet side, where pressurized air can be injected. The side view of the setup (b) shows the high speed camera placed above. The glass plates are clamped together with aluminum framing while the cell gap is controlled with 1 mm thick spacers, which are not shown here.}
\label{fig:1}
\end{figure}

The experimental setup is a linear Hele-Shaw cell, partially filled with Ugelstad spheres \cite{Ugel1980}, i.e. dry, non-expanded polystyrene beads with a diameter of \SI{80}{~\micro\meter} $\pm$ 1 \%. The cell is made out of two rectangular glass plates (80$\times$40$\times$2.5 cm in length, width and thickness respectively) clamped together on top of each other with an aluminum spacer controlled separation of 1 mm. A cell volume (76$\times$32$\times$0.1 cm) is formed between the plates by an impermeable sealing tape as shown in figure \ref{fig:1}, with one of the short sides left open (outlet). Next, beads are filled into the cell by pouring them through the open side until the packing occupies about 90 \% of the cell volume, followed by closing the open side with a semi-permeable filter (a \SI{50}{~\micro\meter} steel mesh) to keep beads inside the cell while allowing air to escape. The cell is then flipped vertically to place the granular medium against the semi-permeable outlet by using gravity, resulting in a volume packing fraction of approximately $\rho_s = 0.44\pm 0.04$, assumed to be more or less uniform across the medium. (We measure the weight and volume of the initial granular layer, and we know the bulk density of the bead material. Then, we calculate the mass by volume for the granular layer and divide it by the bulk density of the beads. The number $\rho_s = 0.44\pm 0.04$ is the average for $\sim$30 experiments $\pm$ the standard deviation.) This leaves a volume empty of beads on the sealed side of the cell, opposite to the semi-permeable outlet, with a linear air-solid interface. An inlet hole on the sealed side of the cell is connected to a pressurized air tank which lets us inject air at a constant and maintained overpressure, $P_{in} = P_{abs,in} - P_0$ (absolute pressure - atmospheric pressure), ranging from 5 to 250 kPa. This will force air to move through the granular medium, towards the semi-permeable outlet, where $P_{abs,out}=P_0=100$ kPa, or in terms of overpressure above the atmospheric one, $P_{out}=0$. The experiments are prepared by the same procedure, resulting in reproducible packing fractions and volume of the initial granular layers. However, there is a randomness in the initial granular configurations, i.e. stress chains and local friction vary from experiment to experiment which introduces some noise to the results. We consider the experiments to be reproducible in the sense that we observe a typical behavior for repeated experiments. In addition, readers may note that the channels tend to grow along the center of the cell. We did not observe a focused displacement of beads in front of the inlet hole as if there was a jet of air initiating the channel growth here, so it appears to be an effect of the confinement rather than the position of the inlet hole, i.e. the beads are easier to displace further from the lateral confining walls.

During experiments, the prepared cell is positioned horizontally. A selected overpressure is set at the pressure tank outlet and verified by a Honeywell pressure sensor with an accuracy of $\pm$4 kPa. The tubing between the pressure tank and the cell inlet is equipped with an electronic valve such that the air injection is started with a digital trigger signal. Positioned above, with a top-down view of the cell, a Photron SA5 high speed camera is started with the same trigger signal, recording the air invasion at a framerate of 1000 images/s and a resolution of 1024$\times$1024 pixels (1 pixel $\approx$ 0.7 mm in the cell). Light from a 400 W Dedolight studio lamp provides uniform and flicker-free illumination onto the white beads of the medium. A small fraction of the beads ($<$ 10 \%) are dyed black with ink to create tracer particles that are used for tracking frame-to-frame deformations in the granular medium. The experiments are run for 10 s, but typically the fracturing and/or compaction of the granular medium takes less than 5 s.

\subsection{\label{sec:level2}Image processing}

In analysis of the images from the high speed camera, we investigate the deformation of the granular medium surrounding the channels formed. We perform image processing with Matlab to obtain quantitatively the information contained in the images.

\begin{figure}
\includegraphics[width=\columnwidth,height=0.9\textheight,keepaspectratio]{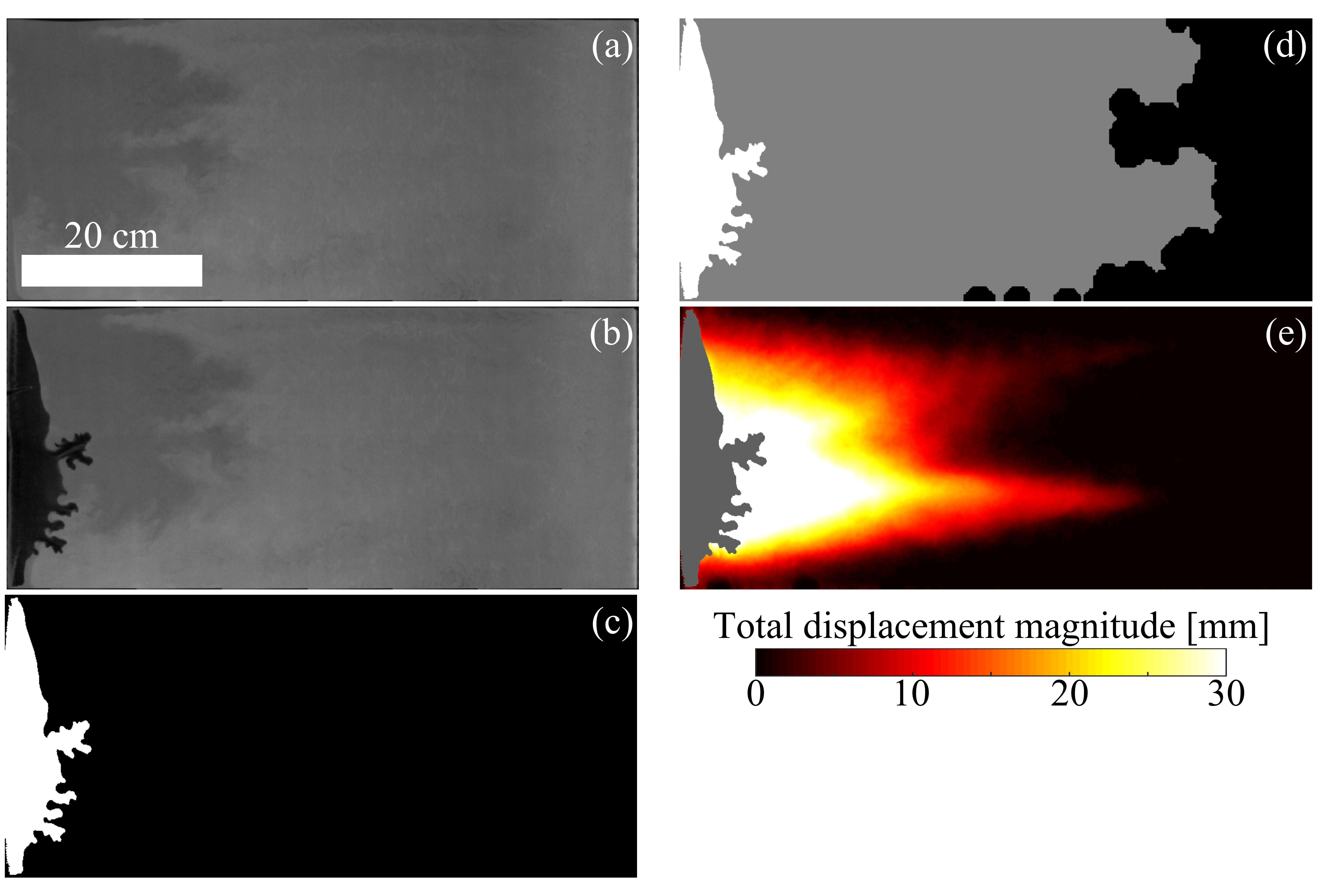}
\caption{Image processing examples for a snapshot in an experiment with injection pressure $P_{in}=200$ kPa. The grayscale snapshot of the initial granular medium (a) ($t=0$) is compared with the grayscale snapshot of the deformed granular medium at the beginning of channel formation (b) ($t=200$ ms), where a region is emptied of beads (black), to produce the binary image (c) where the emptied structure is shown in white. We evaluate deformation with image correlation between frames to find the compacted zone (d) indicated in gray, and the total magnitude of bead displacement as found with DIC (e).}
\label{fig:2}
\end{figure}

The invading structure used as a boundary condition in the pressure simulations is segmented out by converting the grayscale raw data into binary images, where the pixels with value = 1 represent the channel and the pixels with value = 0 represent the granular medium. This is achieved by thresholding each frame in the image sequence with the initial image as reference, such that the pixels having a value less than 30$\%$ of the corresponding initial value are set to 1 (white) and the rest are set to 0 (black), as shown in figure \ref{fig:2} (c).

The frame-to-frame displacement fields are obtained with Ncorr, a Matlab based Digital Image Correlation (DIC) software \cite{Blab2015,BlabWeb}. The basic principle of Ncorr is to cross-correlate subwindows of one image with an image taken at a later time to obtain displacement vectors $\vec{U}(x,y) = u(x,y)\vec{i} + v(x,y)\vec{j}$ located at $(x,y)$ positions in the first image, indicating the displacement of the subwindows between the images.
%
Furthermore, Green-Lagrangian strains are calculated from spatial derivatives of the incremental displacement field as:

\begin{equation}
\begin{split}
E_{xx} &= \frac{1}{2} \left[ 2\frac{\partial u}{\partial x} + \left(\frac{\partial u}{\partial x}\right)^2 + \left(\frac{\partial v}{\partial x}\right)^2  \right]\\
E_{yy} &= \frac{1}{2} \left[ 2\frac{\partial v}{\partial y} + \left(\frac{\partial u}{\partial y}\right)^2 + \left(\frac{\partial v}{\partial y}\right)^2  \right]\\
E_{xy} &= \frac{1}{2} \left[ \frac{\partial u}{\partial y} + \frac{\partial v}{\partial x} + \frac{\partial u}{\partial x}\frac{\partial u}{\partial y} + \frac{\partial v}{\partial x}\frac{\partial v}{\partial y} \right],
\end{split}
\label{eq:strain}
\end{equation}
which for small deformations are similar to small strains: $E_{xx}=\varepsilon_{xx}=\partial u / \partial x$, $E_{yy}=\varepsilon_{yy}=\partial v / \partial y$, and $E_{xy}=\gamma_{xy}=0.5\cdot(\partial u / \partial y + \partial v / \partial x)$, i.e. when the quadratic terms can be neglected. Volumetric strain $\varepsilon_v$ is calculated as the divergence of the displacement field, i.e. $\varepsilon_v = \varepsilon_{xx} + \varepsilon_{yy}$ (assuming that $\varepsilon_{zz} = 0$). Ncorr also includes an algorithm for obtaining the total Lagrangian displacement from the incremental displacement fields. See the article by the developers of Ncorr \cite{Blab2015}, or the web page \cite{BlabWeb}, for an in-depth explanation of the software.

In our analysis, we use subwindows with 20 pixels radius ($\approx$ 14 mm), with their centers separated by a distance of 3 pixels ($\approx$ 2 mm) on a square grid. We use a timestep of 1 ms between successive images, thus obtaining incremental displacements on the smallest timestep possible with our setup. In this paper 5 experiments have been analyzed with DIC, and the method was tested on around 10 experiments where all show the same qualitative results.

We use the total Lagrangian displacements to identify the compacted zone as a region where the total displacement is above a threshold of one tenth of the pixel size. Examples of a compacted zone and displacement field are shown in figure \ref{fig:2} (d) and (e).

\subsection{\label{sec:level2}Numerical simulations}

We use numerical simulations to estimate the evolution of local overpressure values $P(x,y) = P_{abs}(x,y) - P_0$ in the granular medium during experiments. To do this we define a grid with $(I+2)\times(J+2)$ nodes having integer indices $i\in[0,I+1]$ and $j\in[0,J+1]$. By adopting a lattice grid step size $\Delta x=\Delta y=2$ mm, $I$ and $J$ are determined from the length and width of the initial granular medium, respectively $I=700$ mm/$\Delta x=350$ and $J=320$ mm$/\Delta y = 160$.

On this grid, a size-matched binary image (whose dimensions correspond to the initial granular medium) is placed such that all interior nodes with indices $i\in [1,I]$ and $j\in [1,J]$ represent points $(x_i,y_j)$ inside the cell, where $x_i=(i-1/2)\Delta x$ and $y_j=(j-1/2)\Delta y$. The origin $(x=0,y=0)$ is at the lower left corner of the initial air-solid interface at the inlet side. By using these points, we make a discrete representation $P_{i,j}=P(x_i,y_j)$ of the pressure field.

The edge nodes, with indices $i=0$, $j=0$, $i=I+1$ and/or $j=J+1$, represent boundaries around the granular medium. Here, we set fixed boundary conditions, where the pressure at the inlet side is $P_{0,j}=P_{in}$, the pressure at the outlet side is $P_{I+1,j}=P_{out}=0$, and the sealed sides are set to reflect the pressure just inside these boundaries; $P_{i,0}=P_{i,1}$ and $P_{i,J+1}=P_{i,J}$. In addition, the pressure inside the channel empty of beads (found from the binary frames) is held constant at the injection pressure $P_{in}$.

The evolution of air pressure $P$ within the granular medium is given by the equation \cite{Nieb2012a,McNa2000}

\begin{equation}
\frac{\partial P}{\partial t} = D\nabla^2 P - \vec{v}_g\cdot\nabla P - \frac{P}{\phi}\nabla\cdot\vec{v}_g,
\label{eq:airevolution}
\end{equation}
where $\vec{v}_g=v_x\vec{i}+v_y\vec{j}$ is the granular velocity, $\phi$ is the porosity, and $D$ is a diffusion constant as explained below. Equation (\ref{eq:airevolution}) is derived by considering mass conservation of the fluid with a local Darcy law,

\begin{equation}
\frac{\partial (\phi\rho_f)}{\partial t} + \nabla\cdot\left[ \phi \rho_f \left( \vec{v}_g - \frac{\kappa}{\phi \mu}\nabla P \right)\right] = 0,
\label{eq:consflu}
\end{equation}
where $\rho_f$ is the fluid density, $\mu$ is the fluid viscosity and $\kappa$ is the permeability of the medium, combined with mass conservation of the granular medium

\begin{equation}
\frac{\partial \rho_s}{\partial t} + \nabla \cdot\left(\rho_s \vec{v}_g \right)  = \frac{\partial (1-\phi)}{\partial t} + \nabla \cdot\left[(1-\phi) \vec{v}_g \right]  = 0,
\label{eq:consmed}
\end{equation}
where $\rho_s$ is the solid fraction. By using eq. (\ref{eq:consmed}) to eliminate $(\partial \phi/\partial t)$ in eq. (\ref{eq:consflu}), we have

\begin{equation}
\frac{\partial \rho_f}{\partial t} = \frac{1}{\phi \mu}\nabla\cdot(\kappa\rho_f \nabla P) - \vec{v}_g\cdot\nabla\rho_f - \frac{\rho_f}{\phi}\nabla\cdot\vec{v}_g,
\label{eq:densityeq}
\end{equation}
which becomes
\begin{equation}
\frac{\partial P}{\partial t} = \frac{\kappa}{\phi \mu}\nabla\cdot(P \nabla P) - \vec{v}_g\cdot\nabla P - \frac{P}{\phi}\nabla\cdot\vec{v}_g
\label{eq:presseq}
\end{equation}
when assuming that the term in $\nabla \kappa$ is small, and by using the relation $P = \rho_f(P)/(c\rho_f(P_0))$ for an ideal gas, where $c$ is the fluid compressibility. By assuming that there is no strong fluid density variations across the cell (which is valid when the pressure does not vary by orders of magnitude \cite{Turk2015}), we simplify $\nabla(P\nabla P)\approx (1/c)\nabla^2 P$ and arrive at eq. (\ref{eq:airevolution}). The equation basically describes the diffusion of pore pressure in a granular medium, depending on the characteristics of the medium and the interstitial fluid. If the beads move, there is an advective term where the fluid permeates in the moving medium, and where there is divergence in the bead velocity there is a source term for the pressure due to compaction/expansion of the pore volume. See e.g. \cite{Nieb2012a,McNa2000} for a detailed derivation.
The diffusion constant $D$ is defined during the derivation of eq. (\ref{eq:airevolution}) and depends on the compressibility $c$ and the viscosity $\mu$ of the pore fluid, as well as the permeability $\kappa$ and the porosity $\phi$ of the medium (as shown in eq. (\ref{eq:diffconst}) below) \cite{Nieb2012a,Turk2015}. In our system where the granular medium consists of spherical beads and the pore fluid is air, the overpressure diffuses through the pore-space with a diffusion constant given by
\begin{equation}
D=\frac{\kappa}{c \phi \mu} = \frac{\kappa P_0}{\phi \mu}=\frac{d^2\phi^2P_0}{180(1-\phi)^2\mu},
\label{eq:diffconst}
\end{equation}
assuming that the Carman-Kozeny expression is valid for the permeability $\kappa$, and that air is an ideal gas with constant compressibility $c = 1/P_0$. In equation (\ref{eq:diffconst}), $d=$ \SI{80}{~\micro\meter} is the bead diameter, $\mu=1.81\cdot10^{-5}$ Pa$\cdot$s is the viscosity of air, and $P_0=100$ kPa. Due to deformation, the porosity $\phi$ is estimated from the binary image at each timestep; by approximating the solid fraction $\rho_s$ as uniform in the medium for the estimation of the permeability, and assuming that the invading channel is completely empty of beads we get

\begin{equation}
\phi(t)=1-\rho_s(t)=1 - \rho_{s,0}\frac{A_0}{A_0-A_c(t)},
\label{eq:porosity}
\end{equation}
where $\rho_{s,0}=0.44$ is the initial solid fraction, $A_0$ is the initial area of the granular medium and $A_c(t)$ is the channel area as function of time. In our experiments, equation (\ref{eq:porosity}) gives porosities in the range $\phi\in[0.38,0.56]$ (0.56 initially) with corresponding diffusion constants $D\in[7.4\cdot10^4,3.2\cdot10^5]$ mm$^2$/s ($3.2\cdot10^5$ mm$^2$/s initially).

By neglecting the source terms corresponding to the internal deformations (this is discussed in the A1 section of the appendix), i.e. $\vec{v}_g=\vec{0}$, we solve the diffusion equation for the pressure field

\begin{equation}
\frac{\partial P}{\partial t}=D\left(\frac{\partial^2 P}{\partial x^2}+\frac{\partial^2 P}{\partial y^2}\right)
\label{eq:Diffusion}
\end{equation}
numerically during a given timestep with the corresponding diffusion constant by using the Crank-Nicholson scheme \cite{Pres2001}. This is done by solving the set of linear equations for all interior points $i\in[1,I]$, $j\in[1,J]$:

\begin{equation}
\begin{split}
&(1+2\alpha)P_{i,j}^{n+1} \\
&-\frac{\alpha}{2}\left( P_{i-1,j}^{n+1}+P_{i+1,j}^{n+1}+P_{i,j-1}^{n+1}+P_{i,j+1}^{n+1} \right)\\
&= (1-2\alpha)P_{i,j}^{n} \\
&+\frac{\alpha}{2}\left( P_{i-1,j}^{n}+P_{i+1,j}^{n}+P_{i,j-1}^{n}+P_{i,j+1}^{n} \right),
\end{split}
\label{eq:C-N}
\end{equation}
where $\alpha=D\frac{\Delta t}{(\Delta x)^2}=1/2$, giving $\Delta t\in[6.25\cdot10^{-6},2.70\cdot10^{-5}]$ s, and $n\geq 0$ is an integer time index such that $t(n)=n\Delta t$. To obtain the total pressure diffusion during an experiment, we go through the sequence of binary images and let the pressure diffuse for 1 ms per frame (recalling that the framerate is 1000 images/s); when 1 ms is reached, the air cluster and related boundary condition is updated with the next image in the sequence, followed by another 1 ms of diffusion, and so on as a quasi-static evolution (The result of the previous step is used as initial pressure field for the following one). The error sources in this estimation of the pressure field evolution are that we neglect the bead motion and assume a homogeneous porosity throughout the granular medium. However, as mentioned, we show in appendix A1 that the terms containing bead motion during deformation are negligible compared to the diffusion term containing air permeation and pore pressure propagation. To support the approximation of spatially homogeneous porosity, we do not observe a significant change in the results for a rigid medium with a spatially homogeneous porosity that is constant with time and a deforming medium with a spatially heterogeneous porosity that evolves with time during an initial period of the experiments (from $t=0$ ms to around $t=250$ ms, see the initial pressure diffusion curves in fig. \ref{fig:B}). This initial period is typically the time it takes for the compacting zone with heterogeneous porosity to propagate from the inlet side to the outlet side. In the compacted state at later times the porosity is more homogeneous, and varies only close to the most advanced channel tips. Thus, even if the spatially homogeneous porosity is a strong approximation, the initial deformation does not seem to have a significant impact on the pressure field evolution in our system, and the approximation seems reasonable to account for lower porosities in the compacting medium at later times.

In addition, for each frame we calculate the steady-state pressure field by solving the 2-D Laplace equation

\begin{equation}
\frac{\partial^2 P}{\partial x^2}+\frac{\partial^2 P}{\partial y^2}=0.
\label{eq:LP}
\end{equation}

This is done by iteratively relaxing the pressure \cite{Pres2001} at all interior points $i,j$:

\begin{equation}
P_{i,j}^{n}=\frac{P_{i-1,j}^{n-1}+P_{i+1,j}^{n-1}+P_{i,j-1}^{n-1}+P_{i,j+1}^{n-1}}{4},
\label{eq:relax}
\end{equation}
with the same grid and boundary conditions as above, until it converges at the criterion

\begin{equation}
RMSE = \sum\limits_{i=1}^I\sum\limits_{j=1}^J\frac{\sqrt{(P_{i,j}^{n}-P_{i,j}^{n-1})^2}}{I\cdot J} < P_{in}\cdot 10^{-8}.
\label{eq:convergence}
\end{equation}

Note that in equations (\ref{eq:relax}) and (\ref{eq:convergence}), $n$ is the iteration number and not a defined time index.

With solutions from equations (\ref{eq:C-N}) and (\ref{eq:relax}), we obtain the evolution of the pressure field over time, as well as steady-state solutions (solutions of the Laplace equation) at each timestep. The diffusion equation is more accurate to represent the physical pore pressure field in the experiments than the Laplace one initially, but the idea is to compute the Laplace solution to compare if, when and where, and with what accuracy the growth can be compared to a Laplacian growth process.


\section{\label{sec:level1}Results}

As expected, we find the granular medium to exhibit either a solid-like or fluid-like behavior. In the fluid-like regime we observe significant deformation, where beads are displaced by amounts of multiple bead sizes, and that the granular medium has a behavior much like a viscous liquid being invaded by air. For example, the air opens up channels in the medium, and we observe a granular Saffman-Taylor like instability \cite{Saff1958}. However, due to the boundary conditions we always end up with a solid-like medium at the end of an experiment, where there is no apparent deformation or bead displacements, and the air is reaching the cell outlet by seeping through the pore-space. Typically, the deformation process during the experiments can be separated into 3 stages; the initial mobilization of beads without significant channel formation, the channel formation and compaction, and the compacted stick-slip stage. Depending on the injection pressure (e.g. for 250 and 50 kPa respectively), the first stage lasts during the initial 40 - 250 ms, the second stage in the following 250 - 1650 ms and the final stage lasts as long as particle rearrangement is possible, typically a few seconds (2 - 4). Because the system becomes jammed, little happens after 5 s.

\subsection{\label{sec:level2}Pressure evolution}
\begin{figure}
\includegraphics[width=\columnwidth,height=0.9\textheight,keepaspectratio]{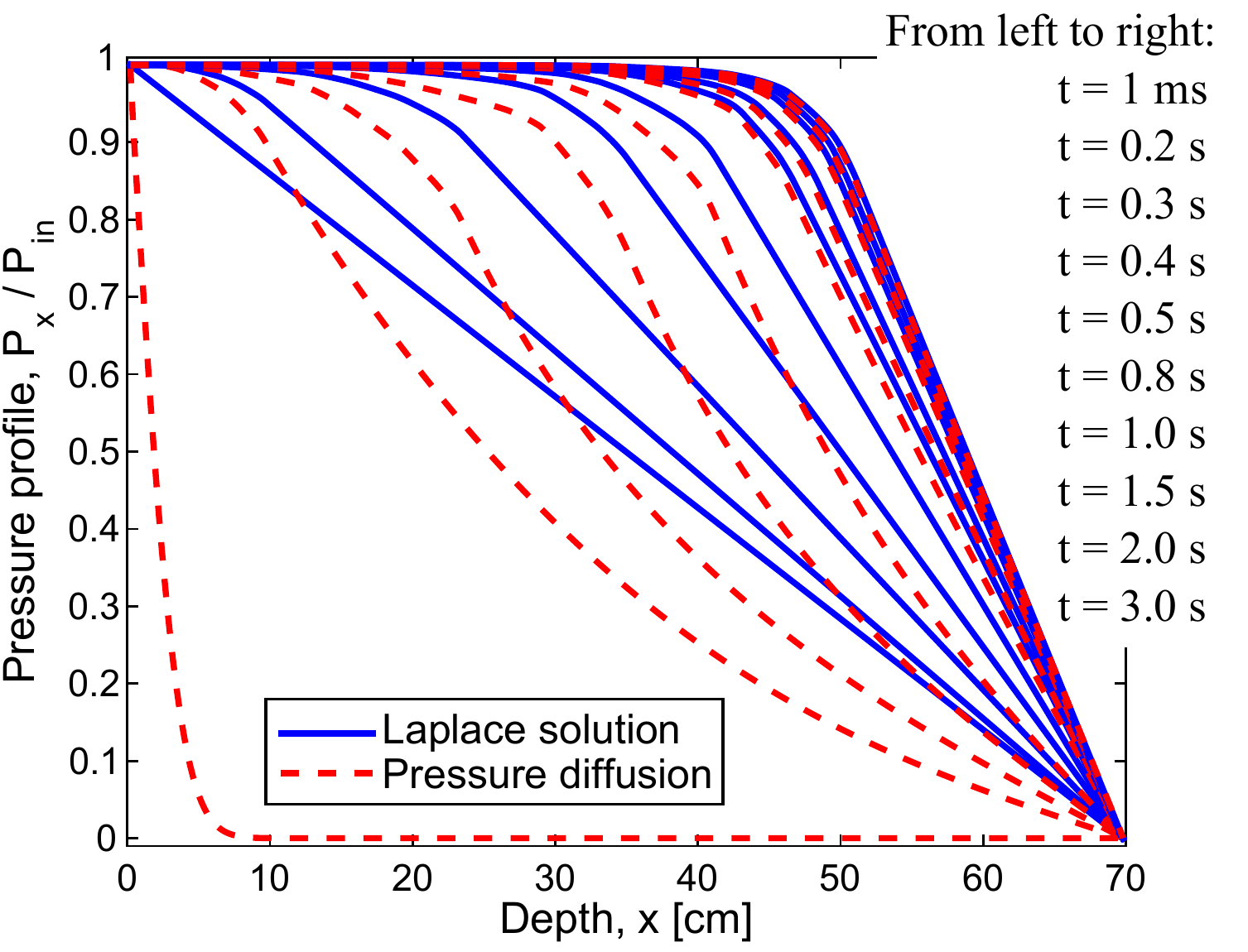}
\caption{Profiles of pressure fields for an experiment with $P_{in}$ = 200 kPa, where the profiles show the average pressure across the cell width (perpendicular to the flow direction). The profiles are the Laplace solution (blue, solid line) and the diffusing pressure (red, dashed line), curves from left to right for each method correspond to the times in the list.}
\label{fig:A}
\end{figure}

Figure \ref{fig:A} shows typical profiles $P_x$ of the simulated pressure field as function of depth into the granular medium, plotted at different times $t$ after the start of injection. The figure shows profiles for both the steady-state Laplace solution and the diffusing pressure field at corresponding snapshots, where the profiles $P_x$ are found as the average pressure across the width of the cell (perpendicular to the flow direction). Early in the experiment, just after opening the overpressure valve ($t=1$ ms), the diffusing pressure field decays quickly as function of depth into the medium, with a range less than 10 cm, while the Laplace solution has a linear decrease of pressure from $P_{in}$ to $P_{out}$ across the cell length (70 cm). Thus, the two simulation methods give quite different solutions initially. Over time, both solutions of the pressure field evolve due to the opening of channels empty of beads, i.e. they move towards the cell outlet as the boundary conditions change. In addition, the profile of the diffusing pressure approaches the profile of the Laplace solution. Thus at later times, e.g. after around t = 0.8 to 1 second, the two methods give practically equal solutions for the pressure fields.

\begin{figure}
\includegraphics[width=\columnwidth,height=0.9\textheight,keepaspectratio]{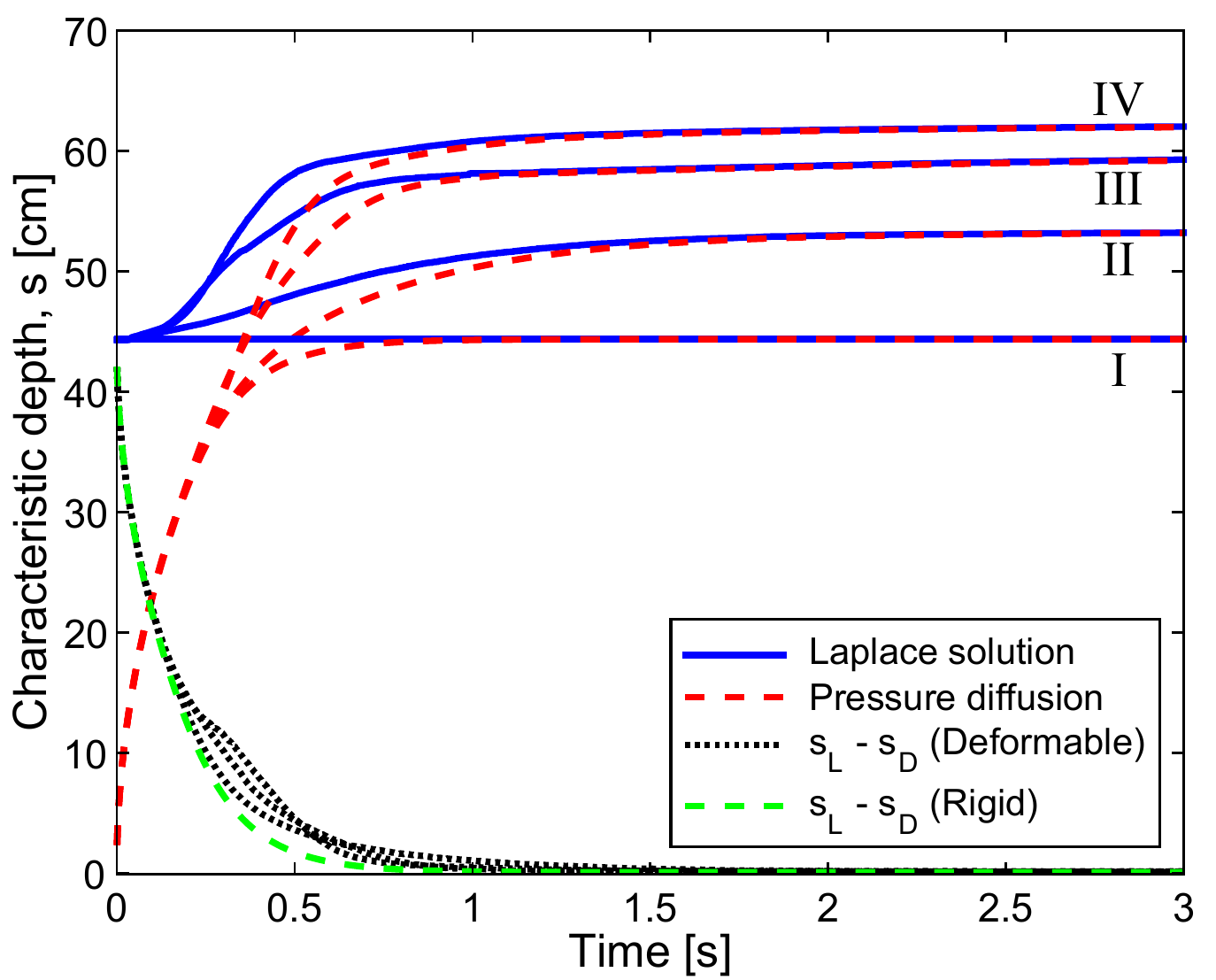}
\caption{Evolution of the characteristic depths $s_L$ for the Laplace solution (blue, solid line) and $s_D$ for the diffusing pressure (red, dashed line) as function of time for 4 different cases: $I$ is a 1-D rigid medium reference solved analytically, $II$ is an experiment with $P_{in}=100$ kPa, $III$ is an experiment with $P_{in}=150$ kPa and $IV$ is an experiment with $P_{in}=200$ kPa. The difference $s_L - s_D$ for the experiments (black, dotted line) all seem to follow the difference $s_L - s_D$ for the 1-D rigid media reference (green, dashed line).}
\label{fig:B}
\end{figure}

To compare the two pressure fields qualitatively, we look at the evolution of a characteristic depth $s$ of the pressure field with time, where $s$ is the depth into the medium where the pressure profile has decayed to $P_x(s) = P_{in}\cdot e^{-1}$. In figure \ref{fig:B} the characteristic depth is plotted as function of time for both the Laplace solution and the diffusing pressure in 4 different cases $I - IV$. Case $I$ is a reference where the granular medium is rigid with $D=3.2\cdot10^5$ mm$^2$/s, solved analytically in 1-D. The cases $II$, $III$ and $IV$ are obtained from pressure profiles of 2-D simulations using experimental data including invasion channels as boundary conditions, with evolving diffusion constant due to porosity change. We calculated the analytical solution for the 1-D diffusing pressure to be

\begin{equation}
\begin{split}
P(x,t)=&P_{in}\left(1-\frac{x}{L}\right) \\
&- \sum_{n=1}^\infty \frac{2P_{in}}{\pi n}\sin\left(\frac{n\pi x}{L}\right) e^{-\frac{n^2\pi^2}{L^2}Dt},
\end{split}
\label{eq:analytic}
\end{equation}
where $L=70$ cm is the system length. We see in eq. (\ref{eq:analytic}) that for increasing time $t$, the diffusing pressure goes towards the steady-state Laplace solution $P(x)=P_{in}\left(1-\frac{x}{L}\right)$. The injection pressure is $P_{in}=$ 100, 150 and 200 kPa in case $II$, $III$ and $IV$ respectively. In addition, in figure \ref{fig:B} we plot the difference $s_L - s_D$ between the characteristic depth of the Laplace solution $s_L$ and the characteristic depth of the diffusing pressure $s_D$. We see that for all experiments, the characteristic depth $s_L$ for the steady-state moves towards the cell outlet due to the channels formed, further and faster with increasing overpressure. However, the difference $s_L - s_D$ decrease similarly with time for all experiments. As mentioned earlier, we also see here that the two simulation methods give practically equal solutions after around $t = 1$ s, which indicates that the diffusing pressure field is relaxed to steady-state on the order of a second, about 0.2 to 0.5 s after the 1-D rigid reference. Since the main evolution of channels and deformation usually occur within the first second of air injection, we use the diffusing pressure solution for the discussions in this article.

\FloatBarrier
\subsection{\label{sec:level2}Initial mobilization and channel formation stages}
\FloatBarrier

\begin{figure*}[t]
\includegraphics[width=\textwidth,height=0.9\textheight,keepaspectratio]{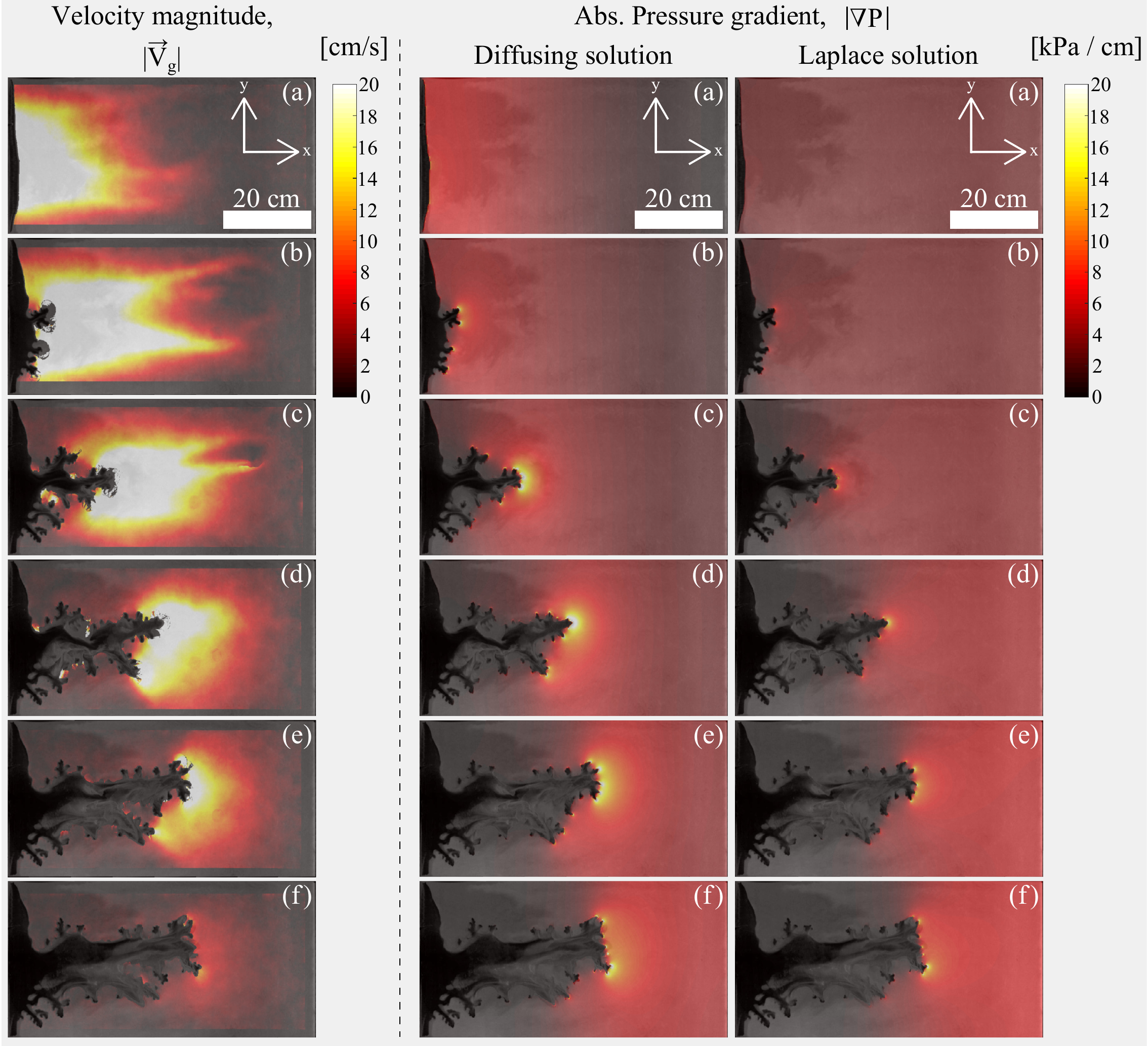}
\caption{Left: Average velocity magnitude $|\vec{v}_g|$ during time windows of $\Delta t = 10$ ms, centered on $t$ = 100, 200, 300, 400, 500 and 600 ms for snapshots (a) - (f) respectively. The injection pressure is $P_{in}=200$ kPa. A zone of mobile beads builds up on the cell scale initially, and later focuses onto the most advanced fingers as the medium compacts. Beads behind the most advanced channel tips are not significantly displaced (Note that we have removed data at the channel, however noise from the erosion inside it appears in (c) - (e)). Right: The absolute pressure gradient $|\nabla P| = \sqrt{(\partial P/\partial x)^2 + (\partial P/\partial y)^2}$ for both solutions of the pressure at the snapshots (a) - (f). The main difference between the diffusing and Laplacian pressure fields is that the diffusing one has higher pressure gradients close to the air-solid interface and finger tips, while it has lower pressure gradients near the outlet boundary. At snapshot (f), the gradients start to look similar in magnitude around the channel. In both solutions, the pressure gradient is screened behind the longest fingers, and the highest magnitudes are on the longest finger tips.}
\label{fig:CD}
\end{figure*}

 The left part of figure \ref{fig:CD} shows snapshots of the magnitude of granular velocity $|\vec{v}_g|$ during 10 ms time windows for an experiment with $P_{in}=200$ kPa. The snapshots are centered on $t = $ 100, 200, 300, 400, 500 and 600 ms, showing typical displacements during the initial mobilization and channeling stages. The first snapshot, at $t=100$ ms is in the later part of the initial mobilization stage. There is still no channels formed, just a slightly curved air-solid interface with a zone of mobilized beads in front of it, spanning about half of the cell. The displacement is mainly in the flow direction, and is higher along the center of the cell and closer to the air-solid interface. The rest of the snapshots, $t=200 - 600$ ms, are taken during the instability and compaction stage. During this stage, fingers open up and form an invasion channel over time. The initially large and spread zone of mobilized beads ahead of the air-solid interface shrinks in size and magnitude over time, focusing onto the tips of the longest fingers. The displacements are largest close to and out from the longest fingers, while there is very little displacement behind the longest fingers.

The right part of figure \ref{fig:CD} shows the absolute values of the pressure gradient $|\nabla P|$ for both the diffusing and Laplacian pressure fields, at the same snapshots as discussed above. Apart from diffusing or being at steady-state, both pressure fields have pressure gradients which are highest on the tips of longest fingers, and has a decreasing magnitude but more evenly distributed across the cell width with distance in front of the longest channel. The pressure gradients are screened behind the most advanced fingers. We see that the most displaced regions in the medium coincide well with the highest pressure gradients, while $\nabla P$ and $|\vec{v}_g|$ goes towards 0 behind the longest fingers. This becomes more evident at later stages when the displacements are more local, as higher pressure gradients are needed to deform the compacted medium further. In figure \ref{fig:CD} (b) and (c) for the granular velocity, the mobilized zones have irregular shapes on the front with relatively large variations in displacement magnitude compared to the corresponding pressure gradients in these regions. This suggests that the displacements in the compacted zone is not only subject to the local pore pressure gradient, but also to solid stresses, i.e. forces related to momentum transfer through bead contacts, as well as local variations in packing density. In \cite{Turq2017a}, numerical simulations include solid stresses and air vibrations to evaluate the total stress state inside the cell.

\begin{figure}
\includegraphics[width=\columnwidth,height=0.9\textheight,keepaspectratio]{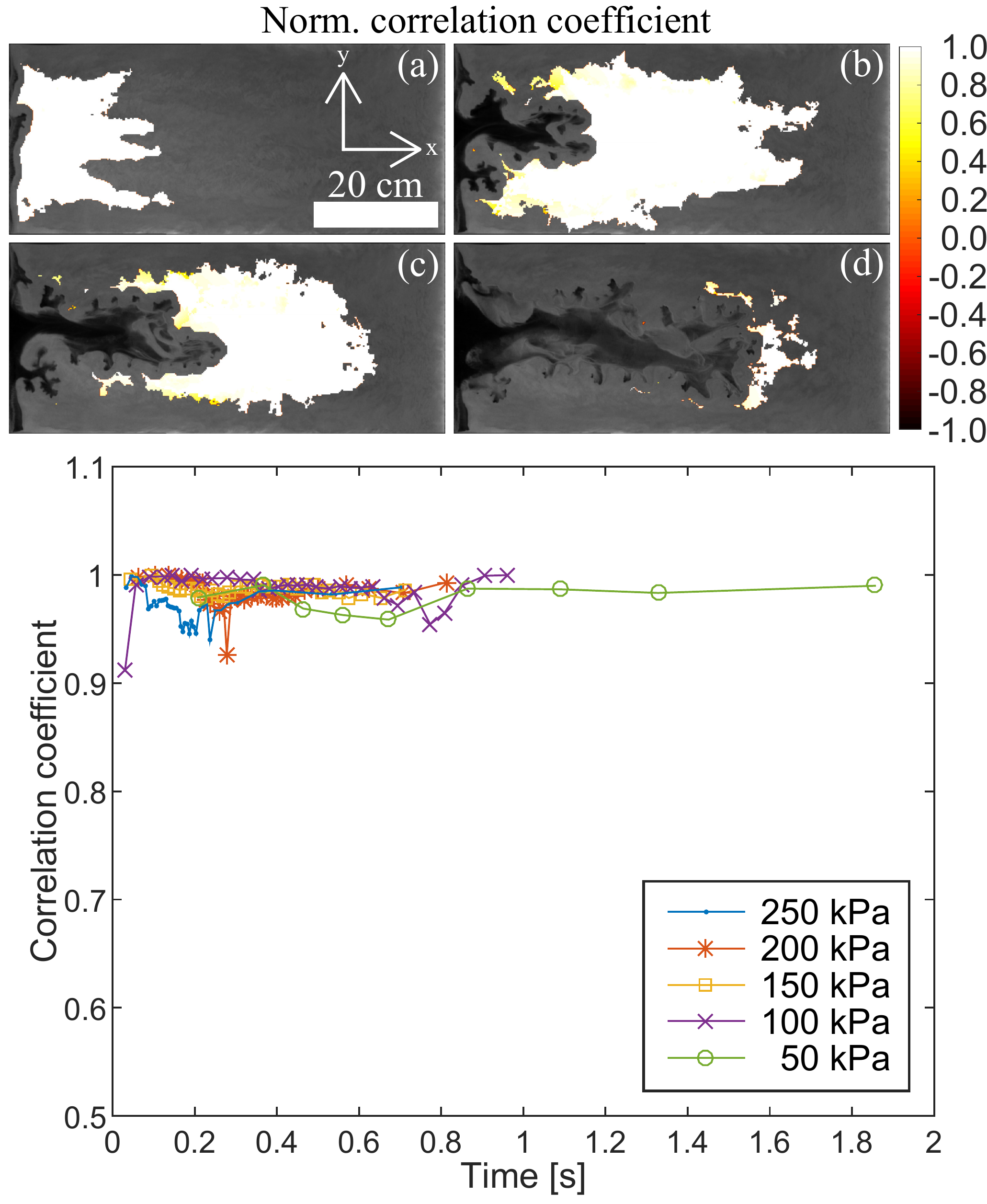}
\caption{Normalized correlation between the vector fields of the negative pressure gradient $-\nabla P$ and granular displacements $\vec{u}_g$ for snapshots separated by 1 ms at different stages of the deformation during an experiment with $P_{in} = 250$ kPa. The data is superimposed on top of the corresponding experimental images during the initial mobilization (a), during the instability stage just after the compaction front has reached the outer boundary (b), during the instability stage with a more compacted medium (c), and in the compacted stage (d). The correlation coefficient is the cosine of the angle between vectors, and is close to 1 in all snapshots, meaning that the beads are generally displaced in the direction against the pressure gradient, i.e. in the direction of the driving force. Displacement magnitudes less than one tenth of a pixel size ($\approx$ \SI{70}{~\micro\meter}) are considered noise and not included (gray). The bottom plot shows the average correlation coefficient in areas with displacement magnitude above the noise level plotted as function of time for the experiments analyzed, showing that the normalized correlation coefficient between $-\nabla P$ and $\vec{u}_g$ is close to 1 over time in all cases.}
\label{fig:G}
\end{figure}

To investigate directional correlation between the driving force ($-\nabla P$) and the bead displacements ($\vec{u}_g$), we calculate the normalized correlation coefficient for the directions of the vectors. The normalized correlation coefficient $C_{dir}$ for the direction of two vectors $\vec{a}$ and $\vec{b}$ is obtained by

\begin{equation}
C_{dir} = \frac{\vec{a}\cdot\vec{b}}{|\vec{a}||\vec{b}|} = \cos\theta,
\label{eq:dotprod}
\end{equation}
which is the normalized dot-product of the vectors. The value of $C_{dir}\in[-1,1]$ gives the cosine of the angle $\theta$ between the vectors, where $C_{dir}=1$ means that they are perfectly aligned, $C_{dir}=0$ that they are perpendicular, and $C_{dir}=-1$ that they have opposite directions. In figure \ref{fig:G} (a)-(d) we have used equation (\ref{eq:dotprod}) to correlate the directions of $-\nabla P$ and $\vec{u}_g$ at each grid point where $|\vec{u}_g|>$ \SI{70}{~\micro\meter} (one tenth of a pixel size) on intervals of 1 ms at different stages during the deformation (Note that the correlation data is superimposed on the experimental images, so the dark areas in \ref{fig:G} (a)-(d) show the granular media and empty channels where we do not have displacement). We see that the correlation is near 1 at all locations and snapshots, suggesting that the beads more or less always move in the direction of $-\nabla P$. To show this fact for all experiments, the bottom part of figure \ref{fig:G} includes a plot of the average $C_{dir}$ (taken over the zones where $|\vec{u}_g|$ is larger than one tenth of the pixel size) as function of time calculated for intervals where the channel length grows by $2.5$ \% increments of the final channel length.

\begin{figure}
\includegraphics[width=\columnwidth,height=0.9\textheight,keepaspectratio]{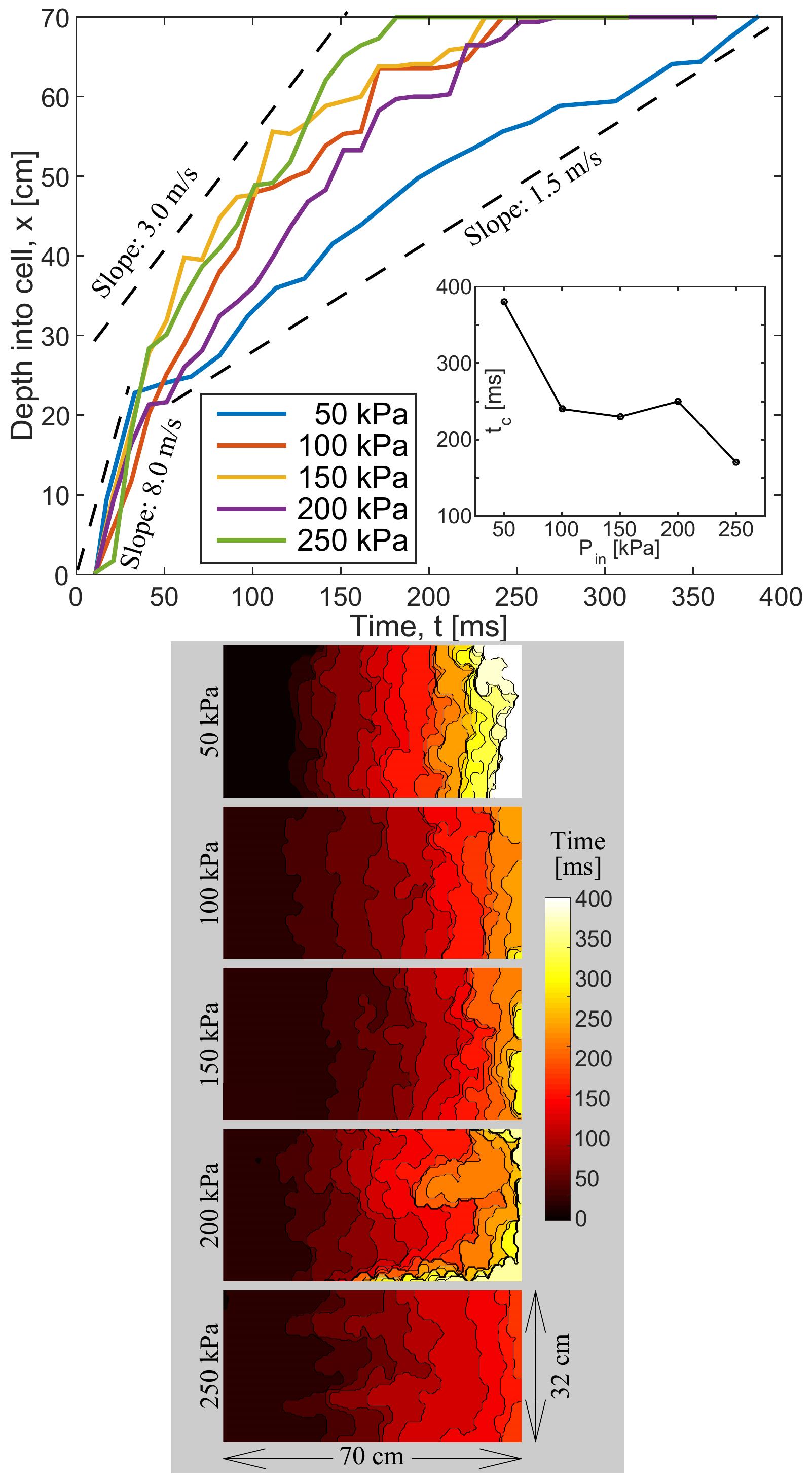}
\caption{Top: Depth of the compacted zones into the medium plotted as function of time for experiments with different injection pressure. The inserted slopes (dashed lines) indicate velocities for reference. Typically, the compacted zone reaches the outer boundary between $t=150$ and 250 ms after the start of injection, however for the lowest pressure it takes around 380 ms. The propagation of the compacted front has a common behavior for all experiments initially. First there is a 10 ms delay after the start of injection until any visible deformation is observed, then the growth is typically faster ($\approx$ 8 m/s) over the first 20 to 30 cm before crossing over to a slower growth rate. The compacted zone in the 50 kPa experiment grows at a rate averaging around 1.5 m/s, while in the 250 kPa experiment the rate fluctuates around 3 m/s. In the 100, 150 and 200 kPa experiments the compacted zones grow at a rate similar to the 250 kPa experiment until a depth of $x$ = 50 to 60 cm, where they slow down, probably due to jamming of the system, but continue to propagate at an average rate around 1.5 m/s as the pressure gradients increase. Inset: Time $t_c$ when the front hits the outlet as function of $P_{in}$. Bottom: Visual representation of the evolution of the compaction zones where the color code represents the time of the snapshot.}
\label{fig:H}
\end{figure}

Figure \ref{fig:H} shows the evolution of the compacted zones with time for experiments with different $P_{in}$. The compacted zones are obtained by thresholding the total displacement fields (from DIC) with a displacement threshold of \SI{70}{~\micro\meter}, i.e. one tenth of a pixel size. The result gives a binary image with white pixels where beads have been significantly displaced from the initial configuration and black pixels otherwise. Then, to reduce noise from the displacement data, the binary images are morphologically closed to connect nearby clusters of white pixels. Here, morphological closing is done as follows: first, all pixels within a distance corresponding to 1 cm from any white pixel is made white (dilation). Next, in the dilated image, all pixels within a distance of 1 cm from any black pixel is made black (erosion). After morphological closing, we keep only the largest cluster of white pixels and consider this as the compacted zone (A figure to illustrate this image processing technique is found in the A2 section of the appendix). The top plot shows the depth $x$ into the medium of the most advanced part of the compacted zones as function of time. For all experiments, it seems to be a common initial behavior for the expansion of the compacted zone: After the start of air injection there is typically a delay of around 10 ms before any visible displacement occurs. Then, over the first 20 to 30 cm into the medium the propagation is typically a bit faster, at a rate of 8 - 10 m/s, before crossing over to a slower growth rate typically between 1.5 to 3 m/s. After crossing over to a slower rate, the compacted zone in the 50 kPa experiment grows at a more or less constant rate fluctuating around 1.5 m/s until it reaches the outlet side. Similarly, the compacted zone in the 250 kPa experiment grows at a more or less constant rate, but at a higher rate around 3 m/s. In the 100, 150 and 200 kPa experiments, the compacted zones has an expansion rate similar to the 250 kPa experiment until they reach a depth of $x$ = 50 - 60 cm into the medium, where they slow down to about 1.5 m/s, probably due to jamming and increased friction.  The bottom part of figure \ref{fig:H} shows a visual representation of the compacted zones over time.

\begin{figure*}[t]
\includegraphics[width=\textwidth,height=0.9\textheight,keepaspectratio]{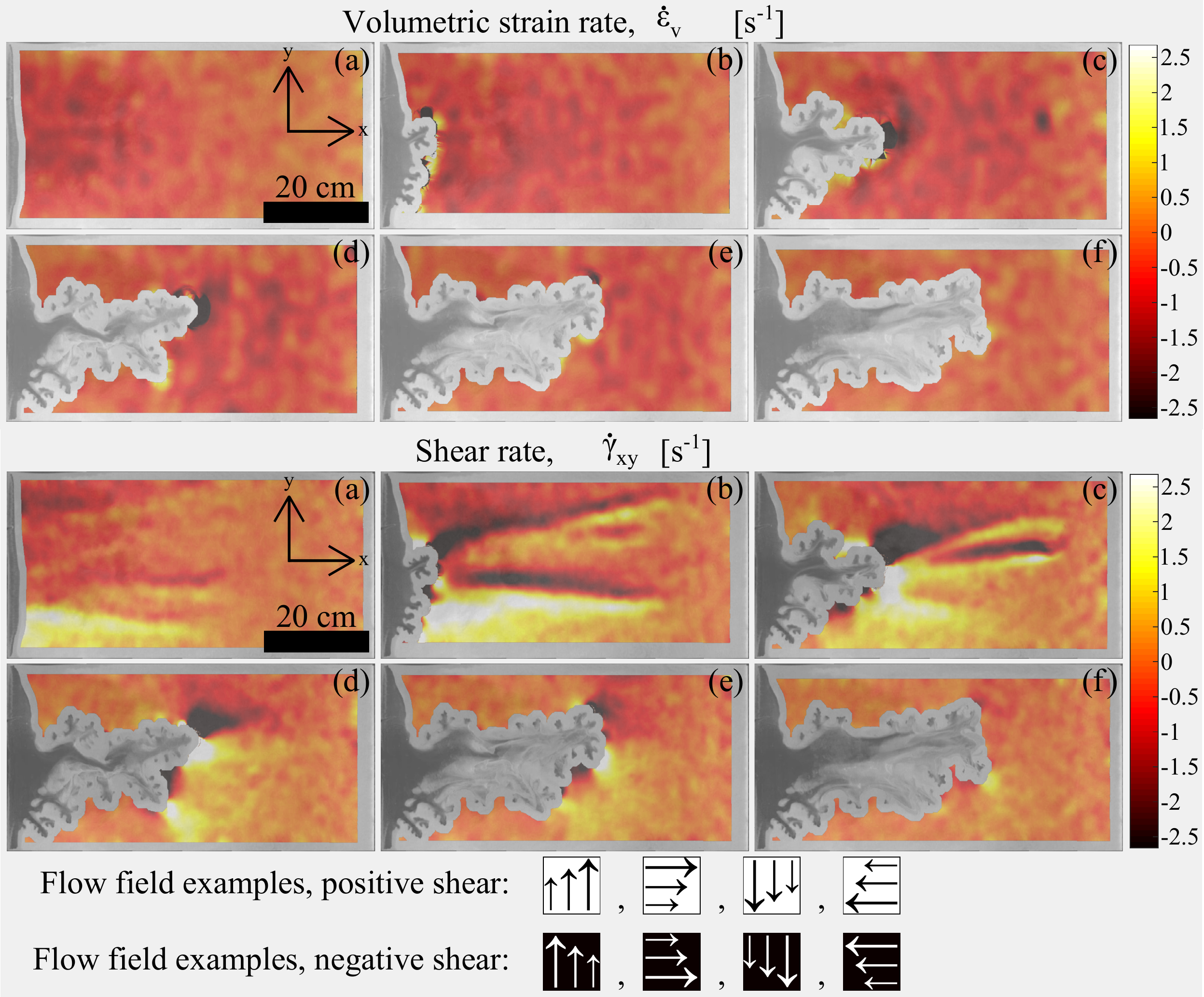}
\caption{Top: Volumetric strain rate $\dot{\varepsilon}_v = \nabla\cdot \vec{v}_g$ in snapshots centered at $t=$ 100, 200, 300, 400, 500 and 600 ms for snapshots (a) - (f) respectively, for an experiment with $P_{in}=200$ kPa. Most of the compaction (negative $\dot{\varepsilon}_v$) occurs in front of the main channel, but some compaction at a lower rate happens on the sides of it. Bottom: Shear rate $\dot{\gamma}_{xy}$ during the snapshots (a) - (f). The sheared regions coincide with the compacting regions, while there is little shear strain behind the longest fingers. We see lines out from the most advanced channel tips separating regions where the shear has opposite signs. These lines can be interpreted as follows; if the shear changes sign from positive to negative across a line (going in the positive $y$-direction), it means that the displacement is higher along the line than the surrounding medium. If the change is from negative to positive, the medium is displaced less along the line than the surrounding medium. The legend on the bottom relates flow behavior with the colors in the deformation map (the shear strain seen here is due to flow from left to right in front of the channel, or up/down on the sides of the main channel).}
\label{fig:EF}
\end{figure*}

The rate of compaction of the medium during experiments is quantified by evaluating the incremental volumetric strain, which is the sum of the incremental normal strains $\varepsilon_{v}=\varepsilon_{xx}+\varepsilon_{yy}$ estimated with DIC from eq. (\ref{eq:strain}) (we assume a 2-D geometry such that $\varepsilon_{zz} = 0$ due to the rigid cell plates). The volumetric strain rate $\dot{\varepsilon}_v$ (or divergence of the granular velocity $\nabla\cdot\vec{v}_g$) is found from incremental volumetric strain as $\varepsilon_v / \Delta t$. Snapshots of the volumetric strain rates in the medium are shown on the top in figure \ref{fig:EF}, again for the same snapshots (a) - (f) as in figure \ref{fig:CD}. The compacting zone (negative $\dot{\varepsilon}_{v}$), grows in the flow direction from the cell inlet towards the outlet side. Typically during channel growth, most of the compaction of the medium occurs in front of the channel. Before the compacted zone hits the outlet side (in (a) and (b)), the strain rate seems similar in the compacting areas, while in (c) and (d) it seems to be higher closer to the channel tip. As the channel growth stops during (e) and (f), the compaction rate decreases. There is also some small zones of decompaction close to the air-solid interface associated with the opening of finger tips, where beads may be pushed away from each other.

The bottom of figure \ref{fig:EF} shows snapshots of the shear rate $\dot{\gamma}_{xy}$ during the flow, found from incremental shear as $\gamma_{xy} / \Delta t$, where $\gamma_{xy}$ corresponds to small strains from eq. (\ref{eq:strain}). The snapshots suggest that the displacements are highest in front of the growing channels and decrease with perpendicular distance away, similar to laminar flow in a pipe where the displacement is highest along the center. This shows that the beads in front of the growing channel are sheared in addition to compacted. During the invasion, we observe that local deformations outside separate branches have the same behavior as the cell scale deformation around the main channel.

\subsection{\label{sec:level2}Compacted stick-slip stage}

\begin{figure*}
\includegraphics[width=\textwidth,height=0.8\textheight,keepaspectratio]{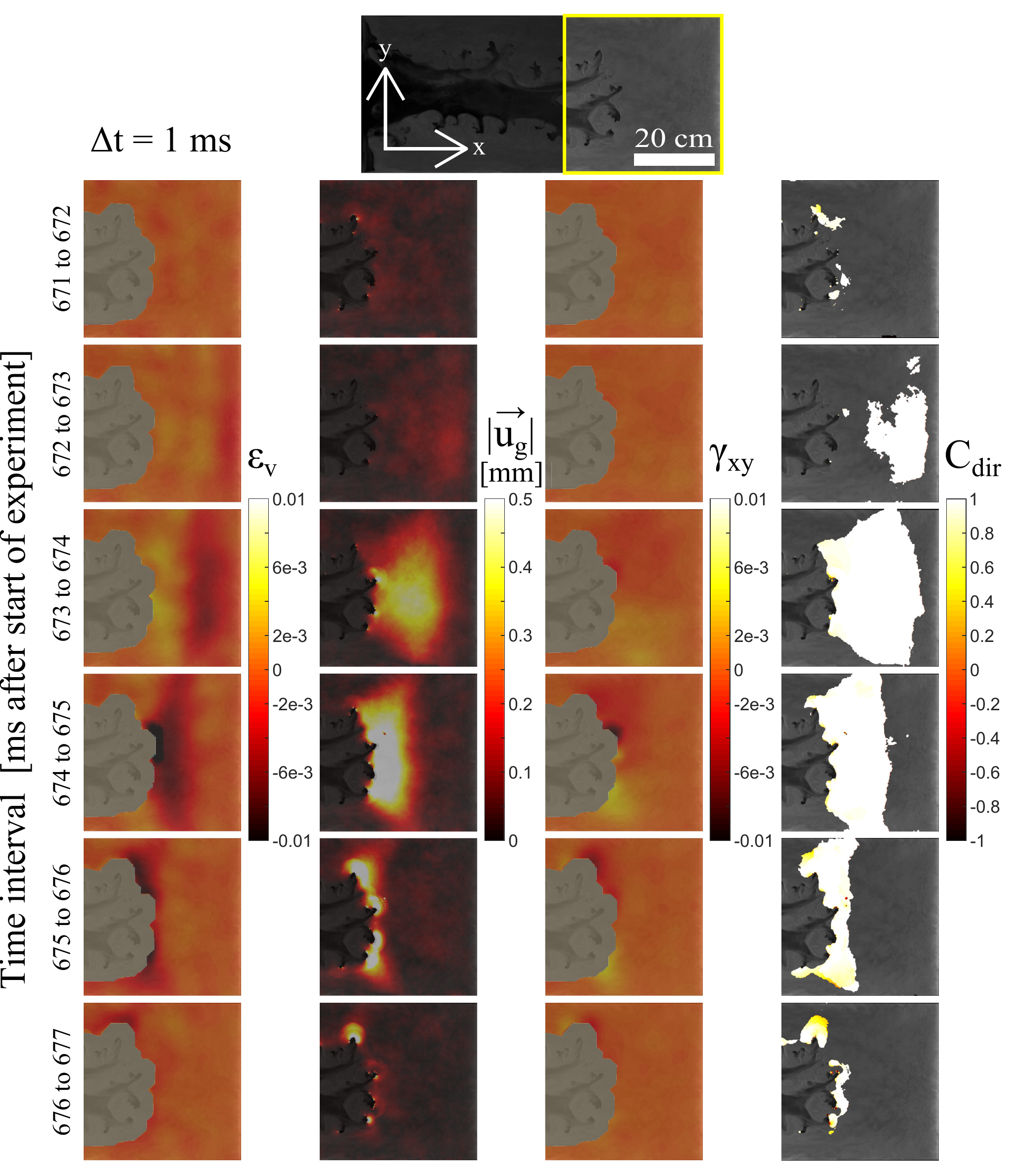}
\caption{Deformations during a stick-slip event in an experiment with $P_{in}=250$ kPa. The snapshots show deformations in a zoomed in region on successive 1 ms intervals, where the time increases from top to bottom. The columns from left to right show: volumetric strain $\varepsilon_v$, displacement magnitude $|\vec{u}_g|$, shear strain $\gamma_{xy}$, and $C_{dir}$, the normalized correlation coefficient between displacements and negative pressure gradient $\vec{u}_g$ and $-\nabla P$. The thumbnail on the top indicates the location of the zoomed in region. A stick-slip event typically lasts between 5 to 10 ms, and the finger tips may expand up to 3 - 5 mm. The deformations shown in the snapshots are typical: Initially there is little deformation (compacted stage) until beads in a region in front of the channel rearrange and compact. Consequently, beads in a zone adjacent to the finger tips de-compact and the fingers propagate as this zone re-compacts, i.e. a de-compaction/compaction front moves from the initially rearranged zone towards the channel and the fingers expand as the front reaches them, such that the closest fingers expand first. The displacements are in the direction of $-\nabla P$, and the medium is also sheared slightly in front of the channel.}
\label{fig:J}
\end{figure*}

The final stage begins when the system is more or less jammed at the cell scale. At this point, the fingers grow slowly and do not deform the medium on a global scale anymore. However, we also observe random and sudden jumps of channel growth, probably due to particle rearrangements in the vicinity of the finger tips. Figure \ref{fig:J} shows a more detailed analysis of the deformations during a stick-slip event in the compacted stage, where 6 successive snapshots of 1 ms intervals display the volumetric strain, displacement magnitude, shear strain, and the correlation between the directions of displacement and opposite pressure gradient during each snapshot. A stick-slip event in this system usually lasts for 5 - 10 ms and the channel tips can propagate up to 3 - 5 mm during this time. The typical steps of a stick-slip event are summarized in the figure: Initially, there is very little displacement since we are in the compacted stage. Suddenly an area ahead of the channel compacts due to rearrangement of beads, causing an area just in front of the channel tips to decompact. The decompacted area re-compacts as a decompaction-compaction front moves from the initially rearranged area towards the channel  - corresponding to a reptation mode of deformation, the sliding zone moving backwards with respect to the grain motion, from the initial nucleation point towards the channel. The channel tips expand quickly when the decompaction-compaction front reaches them, and since the front moves towards the channel, the channel tip closest to the rearranged beads expands before the ones further behind. After this stick-slip event has occurred, the system is back to a jammed state. In addition, we see that the medium is slightly sheared just in front of the channel, and that the displacements are in the direction against the pressure gradient also during stick-slip. This type of event generates intermittent acoustic emissions, as reported in \cite{Turk2015}. This spatio-temporal source shape is characteristic of a seismic pulse \cite{Ampu2008}. Acoustic localization of such events is discussed in \cite{Turq2017b}.

\FloatBarrier
\subsection{\label{sec:level2}Flow in front of channels}

\begin{figure*}
\includegraphics[width=\textwidth,height=0.9\textheight,keepaspectratio]{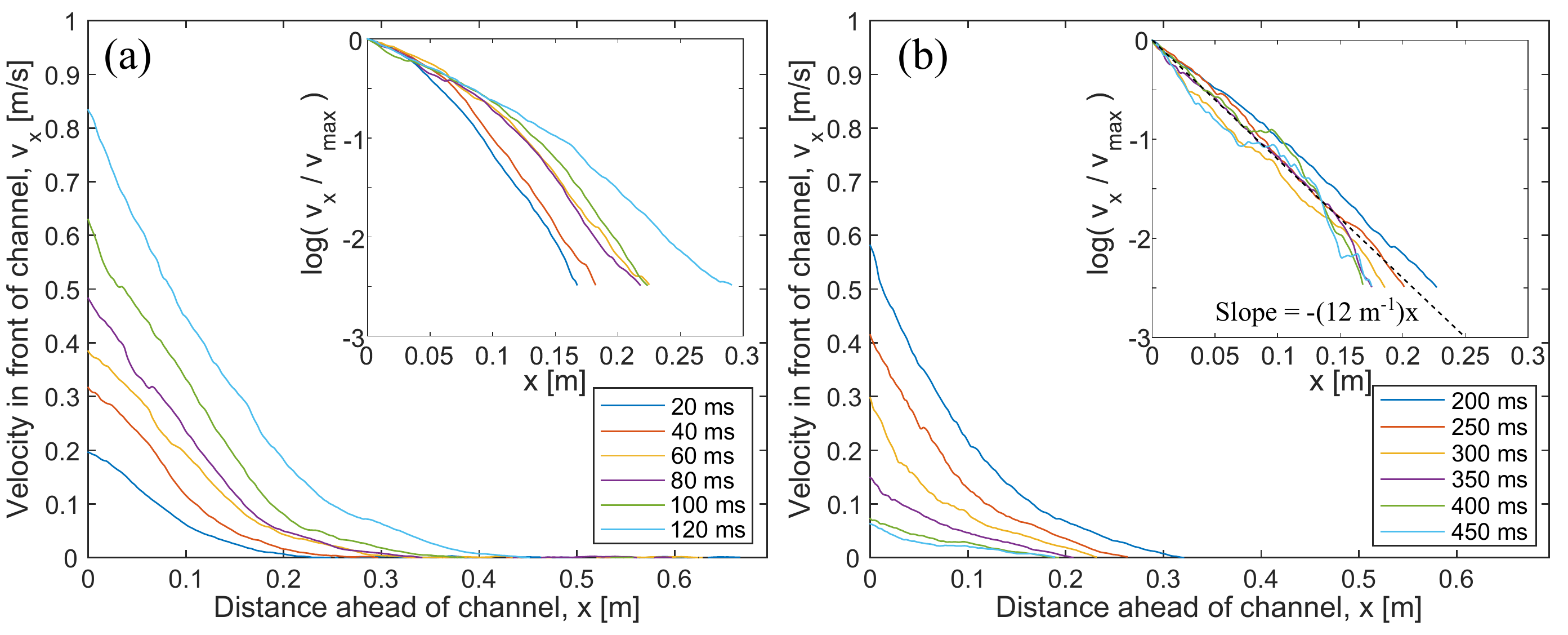}
\caption{Average velocity profiles $v_x$ plotted as function of distance ahead of the channel for different snapshots, where (a) is before the compaction zone reaches the outlet boundary and (b) is after. The insets show the corresponding profiles in semi-log plots. In (a) the decay of $v_x$ with increasing $x$ is more or less linear in the bulk of the compaction zone, while in (b) $v_x$ seems to have an exponential decay.}
\label{fig:frontA}
\end{figure*}

In this section, we present some measurements on the average flow ahead of the growing main channel in an experiment with $P_{in}=250$ kPa, i.e. in a region of interest only covering the length between the channel tip and the outlet side, as well as spanning the central half width of the cell. The average velocity profiles $v_x$ in the flow direction are shown as function of distance ahead of the channel tip for different snapshots in figure \ref{fig:frontA}. The profiles are found as the average $u$-displacement across the width of the region during time windows of $\Delta t=10$ ms, divided by the duration of the time window. The left plot in the figure shows $v_x(x)$ at selected snapshots before the compaction front has reached the outlet side, and the right plot shows $v_x(x)$ at times after the compaction front has reached the outlet side. We see that before the compaction front hits the outlet, the velocity in front of the channel increases with time, and that the velocity has a roughly linear decreasing trend with distance ahead of the channel along the first $\approx 20$ cm of the compaction zone. This suggests that, on average, the compacting strain rate (divergence of $v_x$) is roughly constant with $x$ in the bulk of the compaction zone. The profiles change character after the compaction front has hit the outlet; we now find the velocity in front of the fingers to decrease with time, and that the velocity decreases like an exponential decay with distance ahead of the channel, suggesting that the compaction rate is higher closer to the channel.

\begin{figure}
\includegraphics[width=\columnwidth,height=0.9\textheight,keepaspectratio]{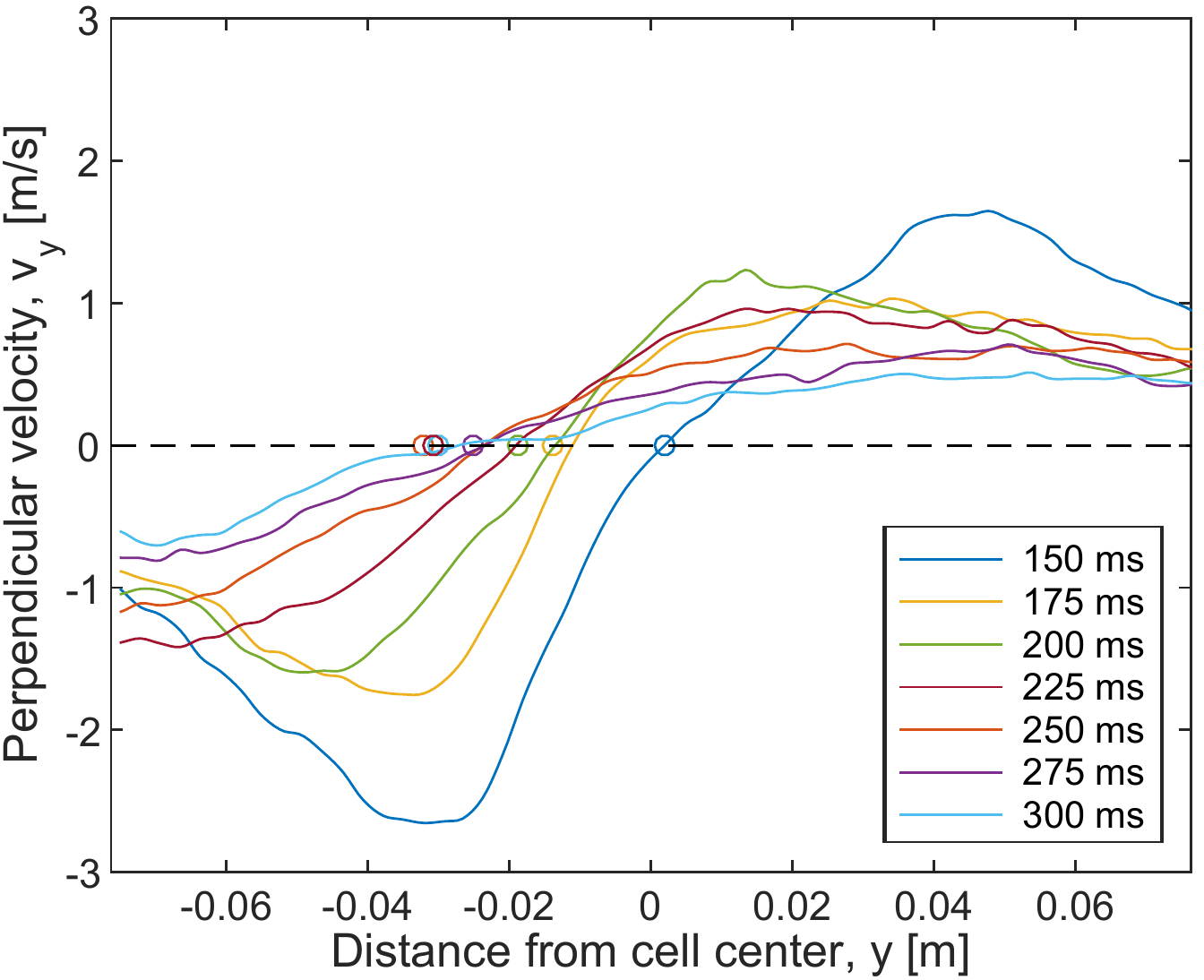}
\caption{Average velocity profiles $v_y$ at 1.5 cm ahead of the channel tip are plotted as function of distance from the center line of the cell, for snapshots during channel growth. The magnitude of the velocity decays with time, but the profiles show that the medium flows away from the channel tip, with positions indicated by open circles.}
\label{fig:frontB}
\end{figure}

The profiles for the velocity $v_y$ perpendicular to the $x-$direction are plotted in figure \ref{fig:frontB} for selected snapshots during channel growth. The profiles show the average velocity along the cell width, in a narrow region of 2.5 cm thickness located $\approx 1.5$ cm ahead of the growing channel. The plot indicates that beads move perpendicularly away from a position close to the most advanced channel tip, with initially increasing velocity as function of distance away from it, indicating decompaction in the $y-$direction close to the tip. Then at positions further away, the medium is compacted. The magnitude of $v_y$ decreases with time as the medium compacts.

\begin{figure}
\includegraphics[width=\columnwidth,height=0.9\textheight,keepaspectratio]{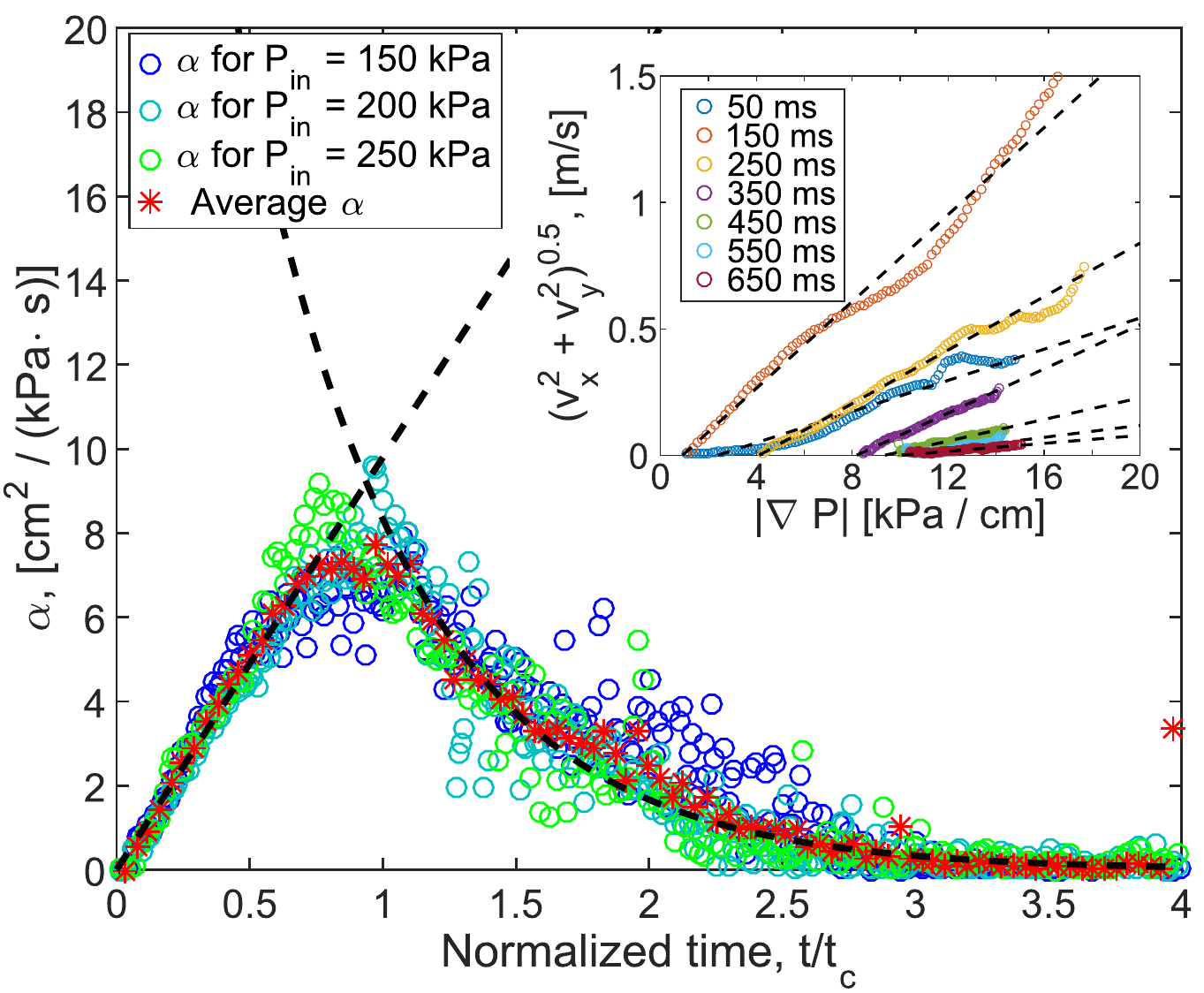}
\caption{Evolution with time of the fit parameter $\alpha$, proportional to the inverse of Bingham effective viscosity of the granular medium, describing the average relationship $|\vec{v}_g|\approx \alpha\cdot |\nabla P| + c$ between the velocity magnitude $|\vec{v}_g|$ and the pressure gradient magnitude $|\nabla P|$. The plots of $\alpha$ are for the experiments with $P_{in}=150$, 200 and 250 kPa, where the data is collapsed along the time axis by $t_c$, the time when the compacted zone reaches the outlet boundary. Before the compaction front reaches the outlet boundary ($t<t_c$), $\alpha$ increase linearly with time (linear fit = $9.8\cdot (t/t_c)$ cm$^2$/(kPa$\cdot$s)), while for $t>t_c$, $\alpha$ has an exponential decay with time (exponential fit = $41.68\cdot e^{-1.61(t/t_c)}$ cm$^2$/(kPa$\cdot$s)). In the inset: Examples of linear fits to the average velocity magnitude as function of pressure gradient. Note that over time there is an increasing offset from the ordinate axis, corresponding to the threshold $\nabla P_c = -c/\alpha$. For the experiments with $P_{in}=$ 150, 200 and 250 kPa, the value of $t_c$ is 230, 250 and 170 ms respectively.}
\label{fig:frontC}
\end{figure}

The velocity magnitude in the considered region of interest (rectangular area between the channel tip and the outlet side spanning the central half width of the cell) is found to have a scattered, but more or less linear relationship with the pressure gradient on average. The coefficient $\alpha$ from linear fits (shown in the inset of figure \ref{fig:frontC})

\begin{equation}
|\vec{v}_g|=\alpha |\nabla P| + c
\label{eq:alphafit}
\end{equation}
describing the average increase of velocity due do an increase of $\nabla P$ is plotted as function of time for the experiments with $P_{in}=$ 150, 200 and 250 kPa in figure \ref{fig:frontC}. To collapse the data along the time axis, we use the normalized time $t'=t/t_c$ where $t_c$ is the time when the compacted zone reaches the outlet side for the respective experiments (found from figure \ref{fig:H}). Respectively, for $P_{in}=50$, 100, 150, 200 and 250 kPa, $t_c=380$, 240, 230, 250 and 170 ms. We see for all experiments that for times before the compacted zone hits the outlet ($t'<1$), $\alpha$ increases linearly with time. Then, after the compaction front has reached the outlet side ($t'>1$), $\alpha$ seems to have an exponential decay with time. The inset in figure \ref{fig:frontC} shows that there is a cutoff threshold $\nabla P_c$ such that $|\vec{v}_g|=0$ if $\nabla P < \nabla P_c$. The cutoff threshold can be found by rewriting equation (\ref{eq:alphafit}) to $|\vec{v}_g|=\alpha (\nabla P - \nabla P_c)$, giving $\nabla P_c = -c/\alpha$. The forces displacing beads in the granular medium are related to the pore pressure gradient  $-\nabla P$, as well as normal and shear solid stresses at contacts between beads. We do not resolve the solid stresses from the experiments (it is however evaluated in \cite{Turq2017a}), but if the main force felt by the medium is due to $-\nabla P$, it suggests a non-Newtonian rheology for the granular medium between the plates. Assuming a Bingham type rheology where
\begin{equation}
\vec{v}_g = -\frac{h^2}{\mu_B}(\nabla P - \nabla P_c),
\label{eq:Bingham}
\end{equation}
and $h$ is the cell gap, the granular paste has an effective viscosity $\mu_B = h^2/\alpha$. 
Then, our results in figure \ref{fig:frontC} suggest that

\begin{equation}
\begin{split}
\mu_B &= \frac{h^2}{\alpha} = \frac{10^{-6}~\textnormal{ m}^2}{9.8~(\textnormal{cm}^2/\textnormal{kPa}\cdot\textnormal{s})}\frac{t_c}{t} \\
&= \frac{10^{-6}~\textnormal{m}^2}{9.8\cdot10^{-7}~(\textnormal{m}^2/\textnormal{Pa}\cdot\textnormal{s})}\frac{t_c}{t} \\
&\sim \frac{t_c}{t}~\textnormal{Pa}\cdot\textnormal{s}
\end{split}
\label{eq:effvisc1}
\end{equation}
before the compacted zone reaches the outlet boundary, where $t_c=380$, 240, 230, 250 and 170 ms for $P_{in}=50$, 100, 150, 200 and 250 kPa respectively (from the inset in figure \ref{fig:H}). After the compacted zone has reached the outlet boundary, the effective viscosity increases as

\begin{equation}
\begin{split}
\mu_B &= \frac{h^2}{\alpha} = \frac{10^{-6}~\textnormal{ m}^2}{41.68~(\textnormal{cm}^2/\textnormal{kPa}\cdot\textnormal{s})}e^{\beta(t/t_c)} \\
&= \frac{10^{-6}~\textnormal{m}^2}{41.68\cdot10^{-7}~(\textnormal{m}^2/\textnormal{Pa}\cdot\textnormal{s})}e^{\beta(t/t_c)} \\
&\sim (0.24~\textnormal{Pa}\cdot\textnormal{s})\cdot e^{\beta(t/t_c)},
\end{split}
\label{eq:effvisc2}
\end{equation}
with $\beta=1.61$ (from figure \ref{fig:frontC}), and $t_c$ from the inset in figure \ref{fig:H}. Thus $|\vec{v}_g|\rightarrow 0$ for $t\gg t_c$, and similarly $|\vec{v}_g|=0$ for small times $t\rightarrow 0$, i.e. the medium is solid-like at these times.

\begin{figure}
\includegraphics[width=\columnwidth,height=0.9\textheight,keepaspectratio]{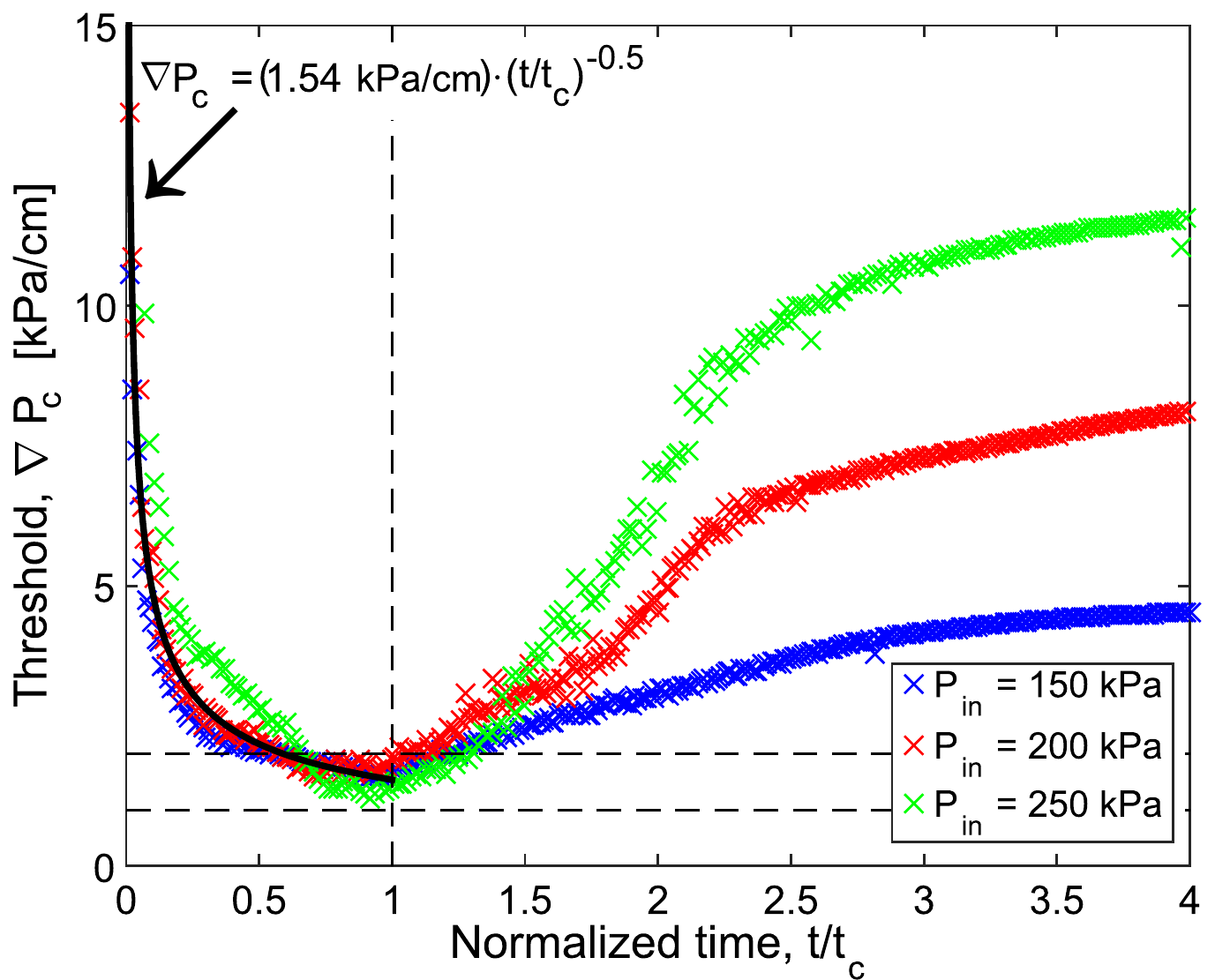}
\caption{Evolution of the thresholds $\nabla P_c$ as function of normalized time $t/t_c$ for the experiments with $P_{in}=$ 150, 200 and 250 kPa, where $t_c=$ 230, 250 and 170 ms respectively. For $t/t_c\leq 1$, the thresholds decrease from 10 - 15 kPa/cm initially to between 1 and 2 kPa/cm (indicated by the horizontal lines) at $t/t_c = 1$. The power law fit (solid black curve) suggests that the decrease in $\nabla P_c$ is inversely proportional to the square root of time. After the compacted zone has reached the outlet boundary, for $t/t_c>1$, $\nabla P_c$ begins to increase with time, faster for higher injection pressure. Then, at $t/t_c$ around 2.5 - 3, the increase of the threshold with time slows down. At this point, $\nabla P_c$ is approaching the pressure gradient values $|\nabla P|$ in the zone surrounding the most advanced finger tips.}
\label{fig:frontD}
\end{figure}

The threshold $\nabla P_c$ evolves during experiments as shown in figure \ref{fig:frontD} for the experiments with $P_{in}=150$, 200 and 250 kPa. The data has been collapsed along the time axis as in figure \ref{fig:frontC}. We see that the average thresholds $\nabla P_c$ decrease similarly before the compacted zone reaches the outlet boundary, decreasing from $\nabla P_c$ between 10 and 15 kPa/cm initially, down to a minimum around 1 - 2 kPa/cm at around the time the compacted zone reaches the outlet boundary. The decrease in $\nabla P_c$ with time seems to fluctuate around the fit

\begin{equation}
\nabla P_c = \frac{\nabla P_{c,min}}{\sqrt{t}}\sqrt{t_c}
\label{eq:thresh}
\end{equation}
as indicated in figure \ref{fig:frontD} for $t\leq t_c$, where $\nabla P_{c,min}$ is around 1.5 kPa/cm. After the compacted zone has hit the outlet boundary, the thresholds begin to increase, faster with higher injection pressure. This probably depends on the injection pressure and the speed of the invading channel, i.e. the compaction rate and the forces available to compact the medium further. As the increase in $\nabla P_c$ slows down (at $t/t_c$ around 2.5 - 3), the thresholds are approaching the pressure gradient in the zone surrounding the most advanced finger tips. Thus, the medium is becoming solid-like and we cross over to the compacted regime. However, the pressure gradients are still 2 - 3 times higher than the thresholds just outside the tips of the most advanced channels, so they propagate slowly. In addition, during the fast channel growth, we can assume that $\nabla P_c$ is small compared to the pressure gradients on the most advanced finger tips, where e.g. $\nabla P$ can be around 40 to 60 kPa/cm for the 250 kPa experiment.

\section{\label{sec:level1}Discussion and conclusion}

In the comparison between the Laplacian pressure field with the diffusing pressure field, we found from figures \ref{fig:A} and \ref{fig:B} that the diffusing pressure reaches the Laplace solution after around 1 s, which is after the main growth of the channels stops. For reference, in the analytical rigid 1-D medium with the same porosity as the initial granular medium, the diffusing pressure reached steady-state after around 0.7 s. These relaxation times can be linked to a characteristic time $\tau$, that depends on properties of the pore fluid, the granular medium and a characteristic length scale $x$, given by the equation

\begin{equation}
\tau = \frac{x^2}{2D}
\label{eq:chartime}
\end{equation}
where $D=\kappa / (c\phi\mu)$ is the pressure field diffusion constant from eq. (\ref{eq:diffconst}). The characteristic time predicts when the effective length of the pressure field reaches a position at a distance $x$ into the medium from the fluid-solid interface. In our system of length $L=700$ mm, $\tau = L^2 / 2D \approx 0.77$ s which is approximately the time it takes for the pressure field to become relaxed, estimated for air and the initial porosity $\phi = 0.56$. By rearranging eq. (\ref{eq:chartime}) to

\begin{equation}
x = \sqrt{2Dt}
\label{eq:charlen}
\end{equation}
we can estimate an effective length $x$ over which the pressure field propagates into the medium from the fluid-solid interface during a given time $t$. For example, in our system with air and porosity $\phi = 0.56$, the pressure field should reach $x \approx 25.3$ mm into the medium when $t = 1$ ms. Thus, characteristic time and length scales can be predicted for the pressure evolution when knowing the properties of the pore fluid and granular medium, by inserting either a relevant length scale into eq. (\ref{eq:chartime}) or a relevant time step into eq. (\ref{eq:charlen}).

During channel growth, the diffusing pressure is similar to the Laplacian pressure in terms of the screening of pressure gradients behind the longest finger tips, as well as having the highest pressure gradients located on the tips of the longest fingers. However, during the fast channel expansion, the magnitude of pressure gradients ($|\nabla P|$) in the diffusing pressure field may be up to 1.5 - 2 times higher than in the Laplace solution in a region close to the most advanced finger tips.

The compacting part of the medium, as shown in figure \ref{fig:H} initially grows in the direction from the inlet towards the outlet, and in all our experiments it reaches the outlet boundary shortly after the start of injection, typically after $t=170$ to 250 ms (380 ms for the 50 kPa experiment). When the compacting zone hits the outlet boundary, the medium more or less instantly becomes harder to displace. We find that the medium exhibits a Bingham like rheology, i.e. it behaves like a non-Newtonian fluid with no deformation at stresses below a yield-stress threshold, while above the threshold the shear rate is proportional to imposed stress minus yield-stress. The constant of proportionality here is considered to be the inverse of the viscosity. In this paper (eq. (\ref{eq:Bingham})) we formulate a similar relationship between deformation and stress in terms of grain velocity, force from the pore-pressure gradient and a threshold pressure gradient $\nabla P_c$. Furthermore, we show in figs. \ref{fig:frontC} and \ref{fig:frontD} that both the viscosity $\mu_B$ and the threshold gradient $\nabla P_c$ evolve during the experiments: Before the compaction front hits the outlet ($t<t_c$), the threshold for bead displacement $\nabla P_c$ appears to decrease with time as the medium mobilizes (note that eq. (\ref{eq:thresh}) is reported more as an observation than a claim), while after the compacted zone hits the outlet ($t>t_c$) $\nabla P_c$ crosses over to increase with time. In a similar fashion, if $-\nabla P$ is the main driving force, the viscosity $\mu_B$ of the medium is suggested to decrease inversely with time before the compacting zone reaches the outlet. After that, it increases exponentially with time due to jamming of the confined system. We interpret the evolution of these material properties by assuming that both $\mu_B$ and $\nabla P_c$ depend on the total solid stress, i.e. a combination of friction (between beads themselves, and with the confining walls) and in-plane solid stress (due to particle contacts). In this case, for $t<t_c$, $\mu_B$ evolves as if the solid stress reduces in the compacting zone as more beads become mobile, and that momentum transfer from the mobile beads lowers the displacement threshold $\nabla P_c$ at the front of the compacting zone. Since friction should increase with the solid fraction $\rho_s$ in the compacting zone, an explanation for the reduced solid stress could be that the mobile beads have a lower dynamic friction coefficient, and as more beads are mobilized, the in-plane solid stress in the compacting zone reduces. At the same time, as more beads become mobile, the compacting zone gains momentum. At later times, for $t>t_c$, the total solid stress builds up fast; The in-plane solid stress builds up due to contact with the outlet boundary, and friction increases as the solid packing fraction continues to increase and beads become immobile again.

The strain rates in figure \ref{fig:EF} show that there is a compacting strain rate across the medium, indicating that the beads are moving faster in the flow direction closer to the invading channel, while the shear rates in the same figure indicate a shear flow where the fastest displacements occur in front of the finger tips, aligned in the growth direction of the fingers. In terms of fracture dynamics for the opening of channels, the deformation shown in figure \ref{fig:EF} suggests that the channel propagates by mode II fracturing with out of plane shear, i.e. shear planes perpendicular to the confining plates, with directions emanating from the most advanced fingers empty of beads. We also expect some in plane shear along the confining plates, assuming that top beads move more easily than bottom beads (not observed by DIC). In other words, the channel is pushing the material in front of it to propagate. At the same time, the typical displacements outside channel tips shown in figure \ref{fig:frontB} indicate a mode I fracture opening, as beads are pushed out to the sides in front of finger tips. Therefore, we suggest that the channels propagate with a mixed mode I/II opening at the tips, i.e. both shear and tensile opening at the same time. This is consistent with the pressure gradients which are pointing radially inwards to the finger tips, and is also supported by findings in \cite{Turq2017b} where the mechanics leading to acoustic emissions are observed (based on the polarisation on the sensors) to have two different types; compaction/relaxation or shearing. 

Our results give insight on how the pore pressure evolves through a quasi-2D granular medium, how the channels open up at the tips as well as the rheology of the granular medium and pore fluid. The main differences between horizontal quasi-2D and 3D systems is that gravity can be neglected and that channels are limited to expand along a given plane in the 2D one. In the translation from quasi-2D to a 3D system, we expect that the evolution of the pressure field would remain universal while the rheology would depend on how deep we look in the medium, i.e. the friction between particles would increase as function of increased overburden, and thus the granular displacement thresholds $\nabla P_c$ and the effective viscosity $\mu_B$ increase with depth. Therefore, we expect for a non-cohesive 3D medium that the growth of eventual channels will be directed upwards if the fluid is injected at a depth (the easiest path), grow along the top layer if the top surface is confined, and also that sub-surface channels may collapse (snap-off) and form isolated pore fluid bubbles. The channel growth mechanism, where beads are pushed in front and to the sides of the channels, is expected to be similar in 3D.
To increase the knowledge about pore fluid overpressure and related deformations in a 3D granular media, it could perhaps be useful to estimate the rheology of a given granular medium and pore fluid in a quasi-2D sample and translate the characteristics into a 3D system by adding gravitational effects. When considering that channels may collapse in 3D granular media, our experiments could translate quite well to pressure evolution and deformation in cohesive 3D porous media where channels remain connected to the inlet. As we showed in figure \ref{fig:J}, it is possible to detect and track the evolution in space and time of the slip velocity field in zones of rearranging beads during stick-slip events lasting less than 10 ms by using DIC. This could prove as a useful tool in the development of acoustic localization techniques by experiments assisted with optical data, where energy release from distinct events of shearing and compacting media can be used to estimate a location for the source region as in \cite{Turk2016}. In such setups it is an advantage to know the kinds of deformations happening, and where they happen, in order to validate measurements and methods. In addition to localization, deformation data can be an aid in discovering mechanisms of characteristic acoustic emissions during channel formation, such as discussed in \cite{Turk2015}. For example, we have observed that the channel propagation typically features small stick-slip events rather than propagating in a smooth movement, especially towards the later part of the compaction stage. Perhaps this can be detected as distinct acoustic events, and even be used for localizing the channel tip during growth.

In the growing zone, where the granular medium is displaced in front of the channel, we found that $v_g\propto (\nabla P - \nabla P_c)$ on average. This is a Laplacian growth $(v\propto \nabla P)$ if $\nabla P_c$ is small, and if the Laplacian field and diffusing field are close to each other locally. It should result in a fractal dimension around 1.71 for the growing channel, similar to DLA clusters \cite{Witt1981} and viscous fingers in empty Hele-Shaw cells \cite{Saff1958}. However, as pointed out in our structural analysis of the pneumatic fractures \cite{Erik2016}, we observe that for low and moderate injection pressures where channels are formed (typically $P_{in} \in [50,150]$ kPa), that the patterns resemble viscous fingers in disordered porous media, while for higher injection pressures ($P_{in} > 200$ kPa) they become more similar to viscous fingers in a classic Hele-Shaw cell. By using box-counting techniques, the fractal dimension was calculated to have typical values between $D = 1.5$ and 1.6 for the low and moderate injection pressures, and typical values between $D = 1.7$ and 1.8 for the higher injection pressures \cite{Erik2016}.

In previous studies on viscous fingers in porous media \cite{Lovo2004,Tous2005}, it was derived that the interface growth should go like $v\propto (\nabla P - \nabla P_c)$, where the thresholds there are due to capillary pressures at the fluid-fluid interface. For moderate capillary numbers and with disorder in the thresholds (i.e. disordered porous media), the authors of \cite{Lovo2004,Tous2005} calculated fractal dimensions with values around 1.5 - 1.6 for the patterns, and derived that the growth in this regime is better described by the Dielectric Breakdown Model (DBM), where $v\propto (\nabla P)^\eta$ with $\eta=2$, rather than DLA where $v\propto \nabla P$. However, for higher capillary numbers where a disorder in the thresholds is less significant, the growth was found to resemble DLA (DBM with $\eta = 1$). In the same sense, considering that we see a Bingham type rheology in our experiments, a disorder in the displacement thresholds could be responsible for the channel growth being in another universality class than DLA for moderate pore pressure gradients, and therefore would be better described by DBM with $\eta=2$. The derivation in Appendix A3 suggests that this could indeed be the case.

\FloatBarrier

\section*{\label{sec:level1}Acknowledgments}

This project has received funding from the European Union’s Seventh Framework Programme for research, technological development and demonstration under grant agreement no. 316889, ITN FlowTrans. We thank Alain Steyer and Miloud Talib for their technical support with experimental equipment, and D. Koehn, V. Vidal, E. Altshuler, A. Lindner, H. Auradou, B. Sandnes, L. Jouniaux, C. Cl\'{e}ment, M. Ayaz and M. Moura for fruitful discussions. This work was partly supported by the Research Council of Norway through its Centres of Excellence funding scheme, project number 262644 (PoreLab), the CNRS LIA France-Norway D-FFRACT, and the INSU ALEAS program.
\vfill
$~$\\[2em]

\bibliographystyle{unsrt}
\bibliography{references}
\clearpage
%
\section*{\label{sec:level1}Appendix}

\subsection*{\label{sec:level2}A1}

Here is a discussion of why we neglect the granular displacements in the simulations of the pore pressure diffusion. When including the granular velocity, the contribution of the second term on the right side of eq. (\ref{eq:airevolution}) in eq. (\ref{eq:C-N}) would be

\begin{equation}
\begin{split}
&\beta_x(P_{i+1,j}^{n+1}-P_{i-1,j}^{n+1}) + \beta_y(P_{i,j+1}^{n+1}-P_{i,j-1}^{n+1}) \\
&\text{ on the left side and} \\
-&\beta_x(P_{i+1,j}^{n}-P_{i-1,j}^{n}) - \beta_y(P_{i,j+1}^{n}-P_{i,j-1}^{n}) \\
&\text{ on the right side,}
\end{split}
\label{eq:2contribution}
\end{equation}
where $\beta_x=\frac{\Delta t}{4\Delta x}(v_x)_{i,j}$ and $\beta_y=\frac{\Delta t}{4\Delta x}(v_y)_{i,j}$. A typical peak in granular velocity in our experiments is $|\vec{v}_g|=1$ m/s, which together with the largest timestep in the simulations $\Delta t = 2.70\cdot10^{-5}$ s gives $\beta_{x,max} = \beta_{y,max} = 3.4\cdot10^{-3} \ll \alpha/2$. Similarly, the contribution of the third term on the right side of eq. (\ref{eq:airevolution}) in eq. (\ref{eq:C-N}) would be

\begin{equation}
\begin{split}
&\gamma P_{i,j}^{n+1} \text{ on the left side and } \\
-&\gamma P_{i,j}^{n} \text{ on the right side,}
\end{split}
\label{eq:3contribution}
\end{equation}
where $\gamma = \frac{\Delta t}{2 \phi}(\dot{\varepsilon}_v)_{i,j}$, and $\dot{\varepsilon}_v=\nabla\cdot\vec{v}_g$. A typical peak in the volumetric strain rate in our experiments is $|\dot{\varepsilon}_v|=2.5$ s$^{-1}$, which together with the largest timestep in the simulations $\Delta t = 2.70\cdot10^{-5}$ s and lowest porosity $\phi = 0.38$ gives $|\gamma_{max}|=8.9\cdot10^{-5}$, which is also negligible. Thus, we approximate the evolution of the pressure field by solving equation (\ref{eq:C-N}), neglecting granular flow and compaction/dilation.

\subsection*{\label{sec:level2}A2}

\begin{figure}[b!]
\includegraphics[width=\columnwidth,height=0.9\textheight,keepaspectratio]{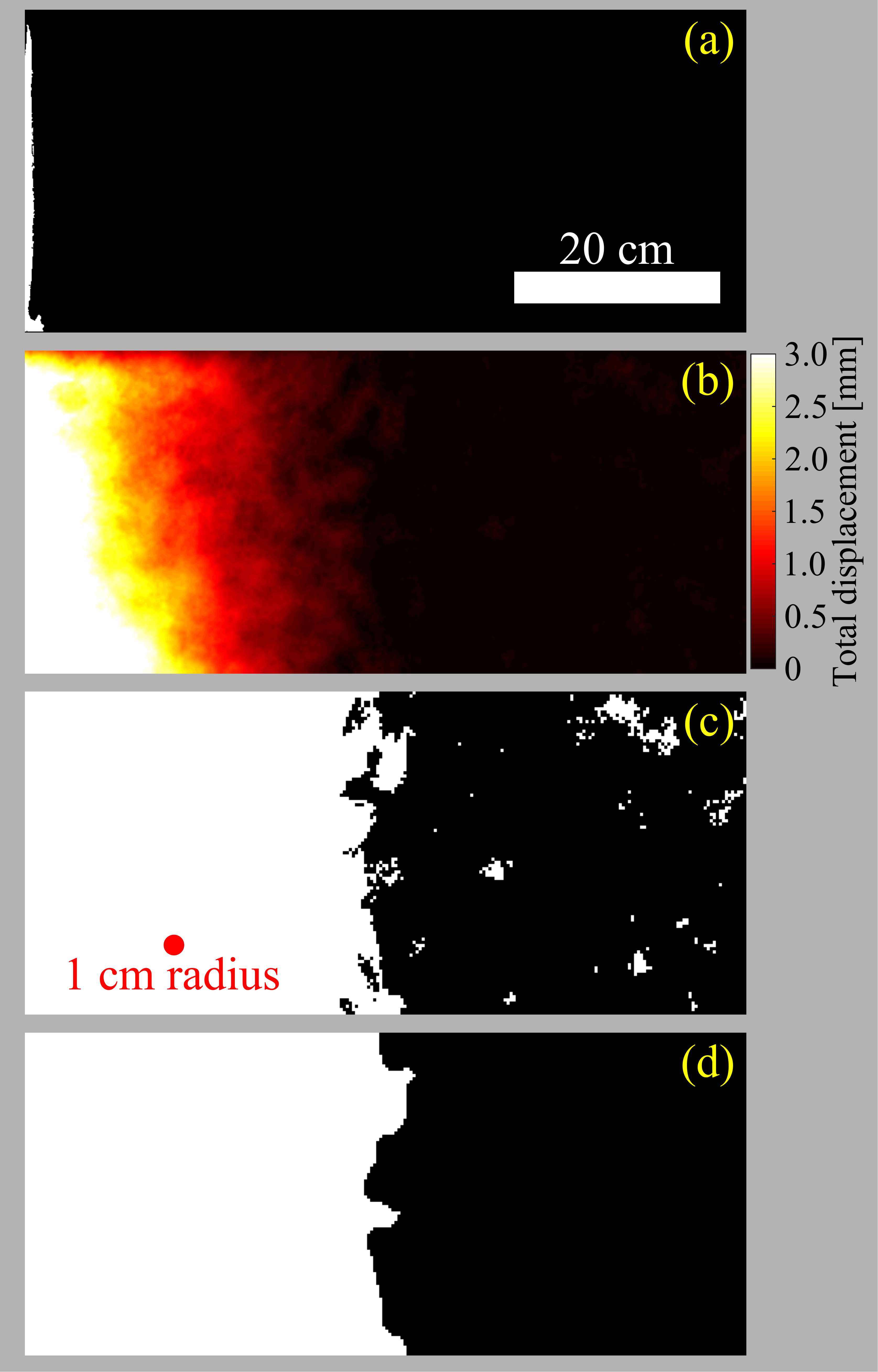}
\caption{The steps to obtain compacted zones from total displacement DIC data. In front of the region invaded by air (a) in a $P_{in} = 100$ kPa experiment at $t = 80$ ms, we obtain the total bead displacement (b) with DIC, which is then converted to a binary image (c) showing the total bead displacements above \SI{70}{~\micro\meter} in white. The smoothened compacted zone (d) is obtained after image treatment of (c).}
\label{fig:cmpexp}
\end{figure}

This section explains the image processing steps to obtain compacted zones from total displacement maps found with DIC. Figure \ref{fig:cmpexp} (a) shows the air invasion (white) in a binary snapshot at $t = 80$ ms in an experiment with $P_{in} = 100$ kPa, and fig. \ref{fig:cmpexp} (b) shows the corresponding total deformation of the granular medium (from DIC). To obtain a binary image of the compacted zone from the total displacement data, we segment the data in fig. \ref{fig:cmpexp} (b) such that all pixels with a displacement more than \SI{70}{~\micro\meter} are assigned the value 1 (white), while the rest are assigned the value 0 (black). The result of this image segmentation is shown in fig. \ref{fig:cmpexp} (c), and we see that in addition to scattered noise the obtained compacted zone has an ill-defined front. Therefore, we smooth the compacted zone by dilating the white pixels with a 1 cm radius disk, followed by eroding the result with the same disk. The dilation step connects white pixels that are spatially separated by distances corresponding to 1 cm or less, while the erosion step approximately restores the initial shape of the white regions. Finally, we remove all but the largest white region to remove the scattered noise. The result is a binary image of the smoothened compacted zone, as shown in fig. \ref{fig:cmpexp} (d). We justify this smoothing technique with the assumptions that beads within the compacted zone are also displaced if they are closer than 1 cm to other deforming beads, and that small deformed regions disconnected from the main compacted zone arise from noise in the DIC calculation (i.e. the scattered noise).

\FloatBarrier
\subsection*{\label{sec:level2}A3}

From equation (\ref{eq:Bingham}) we have the relationship between the pore pressure gradient $\nabla P$ and the grain velocity $v_g$. If we now introduce a disorder in the granular displacement threshold $\nabla P_c$, such that $\nabla P_c \in [\nabla P_{min},\nabla P_{max}]$, we have three scenarios for the interface growth for different pore pressure gradients; first, when $\nabla P < \nabla P_{min}$, $v_g = 0$ and the interface does not expand. Second, when $\nabla P_{min} < \nabla P < \nabla P_{max}$, the driving force is below certain thresholds presented at the interface, i.e. the growth feels the thresholds and is determined by their disorder. Finally, when $\nabla P > \nabla P_{max}$, the driving force is always above the thresholds, and the interface growth is less influenced by their disorder.

We approximate the disorder of $\nabla P_c$ by a flat distribution

\begin{equation}
g(\nabla P_c) = \begin{cases}
    \frac{1}{W}, & \text{for $\nabla P_c \in [\nabla P_{min},\nabla P_{max}]$}\\
    0, & \text{otherwise}
    \end{cases}
\label{eq:DBM_1}
\end{equation}
where $W = \nabla P_{max} - \nabla P_{min}$ is the width of $g$.

We assume that the interface has an average growth velocity independent of the particular arrangement of the thresholds, and find this by averaging over all possible $\nabla P_c$ for a given pore pressure gradient,

\begin{equation}
\langle v_g \rangle = -\frac{h^2}{\mu_B} \int{(\nabla P - \nabla P_c)\theta(\nabla P_c)g(\nabla P_c) d\nabla P_c},
\label{eq:DBM_2}
\end{equation}
where $\theta(\nabla P_c)$ is a Heaviside function which is 0 when $\nabla P \leq \nabla P_c$, and 1 otherwise.
%

In the case where the interface growth feels the disorder in the thresholds, i.e. when $\nabla P_{min} < \nabla P < \nabla P_{max}$, we integrate equation (\ref{eq:DBM_2}) from $\nabla P_{min}$ to $\nabla P$ and get

\begin{equation}
\begin{split}
\langle v_g \rangle &= -\frac{h^2}{\mu_B W} \left[ \frac{1}{2}(\nabla P)^2 - \nabla P \cdot \nabla P_{min} \right.\\ &\left.\qquad\qquad\quad + \frac{1}{2}(\nabla P_{min})^2 \right] \\
&= -\frac{h^2}{2 \mu_B W} (\nabla P - \nabla P_{min})^2.
\end{split}
\label{eq:DBM_4}
\end{equation}

In the case where $\nabla P > \nabla P_{max}$, we integrate equation (\ref{eq:DBM_2}) from $\nabla P_{min}$ to $\nabla P_{max}$ and get

\begin{equation}
\begin{split}
\langle v_g \rangle
&= -\frac{h^2}{\mu_B W} \left[ \nabla P \cdot \nabla P_{max} - \frac{1}{2}(\nabla P_{max})^2 \right. \\
&\left.\qquad\qquad\quad- \nabla P \cdot \nabla P_{min} + \frac{1}{2}(\nabla P_{min})^2 \right] \\
&= -\frac{h^2}{\mu_B W}(\nabla P_{max} - \nabla P_{min})~\cdot \\
&\qquad\qquad\quad\left[ \nabla P - \frac{1}{2}(\nabla P_{max} + \nabla P_{min}) \right] \\
&= -\frac{h^2}{\mu_B}\left( \nabla P - \frac{\nabla P_{max} + \nabla P_{min}}{2} \right),
\end{split}
\label{eq:DBM_5}
\end{equation}
which we see is similar to equation (\ref{eq:Bingham}) when noting that the last term is simply the average threshold value $\overline{\nabla P_c}$. Equation (\ref{eq:DBM_4}) indicates that if there is a disorder in the granular displacement thresholds, $\nabla P_{min}$ is small, and the driving force is within the range of the threshold values, we have a quadratic relationship for the average interface velocity as

\begin{equation}
\langle v_g \rangle \sim (\nabla P)^2,
\label{eq:DBM_6}
\end{equation}
i.e. DBM with $\eta = 2$. However, if the driving force is much higher than the maximum displacement threshold, $\nabla P \gg \nabla P_{max}$, we have instead a linear relationship between the pore pressure gradient and the interface velocity as

\begin{equation}
\langle v_g \rangle \sim \nabla P,
\label{eq:DBM_7}
\end{equation}
i.e. DBM with $\eta = 1$, or DLA.

\end{document}